\pdfoutput=1
\documentclass[graybox, envcountchap]{svmult}

\usepackage{mathptmx}        
\usepackage{amsmath}
\usepackage{amssymb}
\usepackage{color}
\usepackage{helvet}          
\usepackage{courier}         
\usepackage{dirtree}

\usepackage{makeidx}        
\usepackage{graphicx}        
\usepackage{subfig}

\usepackage{multicol}        
\usepackage[bottom]{footmisc}

\usepackage{hyperref}        
\hypersetup{colorlinks=true,urlcolor=blue}

\usepackage[misc]{ifsym}

\newcommand{\beq}{\begin{equation}}
\newcommand{\beqa}{\begin{eqnarray}}
\newcommand{\eeq}{\end{equation}}
\newcommand{\eeqa}{\end{eqnarray}}

\newcommand{\simgt}{\lower.5ex\hbox{$\; \buildrel > \over \sim \;$}}
\newcommand{\simlt}{\lower.5ex\hbox{$\; \buildrel < \over \sim \;$}}

\makeindex             

\begin{document}


\title{Machine-learning applications for weak-lensing cosmology}
\author{Masato Shirasaki}
\institute{Masato Shirasaki (\Letter) \at 
National Astronomical Observatory of Japan (NAOJ), National Institutes of Natural Sciences, 
Tokyo 181-8588, Japan \email{masato.shirasaki@nao.ac.jp},\\
The Institute of Statistical Mathematics, Tokyo 190-8562, Japan,\\
RIKEN Center for Advanced Intelligence Project, 
Tokyo 103-0027, Japan
}
%
%
\maketitle

\abstract{
This chapter reviews recent advances in the application of machine learning to weak-lensing cosmology. Weak gravitational lensing provides a unique and powerful probe of the total matter distribution in the Universe, independent of its physical state. By directly tracing the spatial distribution of otherwise invisible dark matter within the cosmic web, weak lensing has become a cornerstone for studying both the nature of dark matter and the physics governing large-scale structure formation. We begin by introducing the conventional estimators used to extract weak-lensing signals from modern galaxy-imaging surveys and by summarizing established methods for deriving cosmological information from these observables. We then discuss the limitations inherent in traditional analyses and outline how machine-learning techniques can mitigate these challenges. Finally, we explore future prospects for machine-learning-based approaches, highlighting their potential to further enhance the scientific return of current and upcoming weak-lensing datasets.
}

\section{Introduction}
\label{sec:intro}
General relativity predicts that the apparent image of a distant astrophysical source is distorted by gravitational lensing. This distortion is governed by the gravitational potential of intervening matter (the lens) and by the geometric configuration between the source, the lens, and the observer. By measuring coherent distortions in the shapes of a large number of background sources, one can infer the large-scale distribution of gravitating matter in the Universe in an essentially unbiased manner. Lensing distortions induced by the large-scale structure are commonly referred to as weak gravitational lensing, and the resulting reconstructions of the projected mass distribution are known as weak-lensing mass maps. Statistical analyses of such mass maps provide crucial insights into the gravitational clustering of dark matter and the expansion history of the Universe.

This section reviews methods for extracting cosmological information from weak-lensing datasets. We begin with conventional two-point correlation analyses and then introduce a variety of summary statistics designed to capture non-Gaussian information that cannot be fully characterized by two-point correlations alone. We next discuss key limitations of traditional statistical approaches and motivate the need for more flexible, data-driven methodologies. Finally, we provide an overview of roughly a decade of advances in applying machine-learning techniques to weak-lensing cosmology and outline prospects for future developments in this rapidly evolving field.

\section{Preliminaries}
\label{sec:pre}
\subsection{Cosmological parameters}

The matter distribution in the Universe exhibits highly nonlinear and hierarchical structures on small scales, while it is well described as homogeneous and isotropic on sufficiently large scales. This assumption, known as the cosmological principle, underlies the Friedmann–Lema$\hat{\rm i}$tre–Robertson–Walker (FLRW) metric,
\beqa
{\rm d}s^2 = -c^{2}{\rm d}t^{2} + a^2(t)\left[\frac{{\rm d}r^2}{1-Kr^2}+ r^2 ({\rm d}\theta^{2}+\sin^{2} \theta {\rm d} \phi ^{2})\right],
\label{eq.RWmetric}
\eeqa
where $a(t)$ is the scale factor. Throughout this chapter, we assume a spatially flat universe ($K=0$) and normalize the scale factor as $a=1$ today.

The expansion of the Universe is governed by the Friedmann equations, which relate the Hubble parameter $H=\dot{a}/a$ to the total energy density of matter, radiation, and dark energy (DE). Their evolution is conveniently expressed in terms of the density parameters $\Omega_\alpha=\rho_\alpha/\rho_c$, where $\rho_c=3H^2/(8\pi G)$ is the critical density. The resulting expansion history can be written as
\beqa
H^2(a)=H_0^2\left[
\frac{\Omega_{\rm m0}}{a^3}
+\frac{\Omega_{\gamma 0}}{a^4}
+\Omega_{\mathrm{DE}, 0}\exp\left\{-3\int\frac{{\rm d}a'}{a'}[1+w_{\rm DE}(a')]\right\}
\right],
\label{Hubble}
\eeqa
showing that cosmological evolution is fully specified once the density parameters are given.

Structure formation arises from small primordial density fluctuations observed in the cosmic microwave background, which grow under gravity. In the linear regime, the growth of matter overdensities is characterized by a growth factor and statistically described by the matter power spectrum $P_m(k)$. 
The present-day normalization of $P_m(k)$ is commonly quantified by $\sigma_8$, the rms fluctuation amplitude smoothed on $8\, h^{-1}\mathrm{Mpc}$ scales.

\subsection{Weak gravitational lensing}

Weak gravitational lensing refers to small but coherent distortions of background galaxy images caused by the gravitational potential of intervening large-scale structure. These distortions encode direct information about the projected matter density, independent of the physical state of matter, making weak lensing a powerful cosmological probe.

In the weak-lensing regime, the mapping between unlensed and observed galaxy positions leads to an image distortion matrix
\beqa
A_{ij} =
\left(
\begin{array}{cc}
1-\kappa-\gamma_1 & -\gamma_2 \\
-\gamma_2 & 1-\kappa+\gamma_1
\end{array}
\right),
\label{distortion_tensor}
\eeqa
where $\kappa$ is the convergence, describing isotropic magnification, and $\gamma=(\gamma_1,\gamma_2)$ is the shear, describing anisotropic distortions. The convergence is related to the underlying matter density contrast $\delta$ through a line-of-sight projection,
\beqa
\kappa = \frac{3}{2}\left(\frac{H_0}{c}\right)^2 \Omega_{\rm m0}
\int_0^{r}{\rm d}r' \, \frac{r'(r-r')}{r}\frac{\delta}{a},
\label{kappa_delta}
\eeqa
establishing a direct link between weak-lensing observables and cosmological parameters.

In practice, weak lensing is measured from galaxy shapes. The observed ellipticity $\boldsymbol{\epsilon}$ is related to the intrinsic ellipticity $\boldsymbol{\epsilon}_{\rm int}$ and the reduced shear $\boldsymbol{g}$ as \cite{Seitz:1996vf}
\beqa
\boldsymbol{\epsilon}
= \frac{\boldsymbol{\epsilon}_{\rm int} + \boldsymbol{g}}
{1 + \boldsymbol{g}^{*}\boldsymbol{\epsilon}_{\rm int}},
\label{eq:eps_obs_g_eps_int}
\eeqa
which reduces in the weak-lensing limit to
\beqa
\boldsymbol{\epsilon} \simeq \boldsymbol{\gamma} + \boldsymbol{\epsilon}_{\rm int}.
\label{eq:eps_gamma_in_WL}
\eeqa
Since intrinsic ellipticities have a dispersion $\sigma_{\rm int}\simeq 0.4$ \cite{Li:2021mvq}, much larger than the shear signal of individual galaxies, weak-lensing analyses rely on statistical averaging over large galaxy samples.

\subsection{Two-point correlations}

The primary summary statistic in weak-lensing cosmology is the two-point correlation function of shear or, equivalently, the convergence power spectrum. Accounting for the redshift distribution of source galaxies, the convergence can be written as
\beqa
\kappa(\boldsymbol{\theta})
&=& \int_0^{r_H} {\rm d}r\, W_{\kappa}(r)\, \delta(r, r\boldsymbol{\theta}), \\
\label{kappa_delta_v2}
W_{\kappa}(r) &=& \frac{3}{2}\left(\frac{H_0}{c}\right)^2 \Omega_{\rm m0}
r \int_{r}^{r_H}{\rm d}r_s \, p(r_s) \, \frac{r_s-r}{r_s},
\eeqa
where $r_H$ is the comoving distance to the horizon and $p(r)$ 
is the source galaxy distribution, which is normalized as $\int \mathrm{d}r \, p(r) = 1$.
This leads, under the Limber approximation \cite{Limber:1954zz}, to
\beqa
P_\kappa(\ell)=\int_0^{r_H} {\rm d}r\, \frac{W_\kappa^2(r)}{r^2}
P_{m}\left(k=\frac{\ell}{r}, z(r)\right).
\label{eq:pow_kappa}
\eeqa

Transforming back to configuration space yields the commonly used shear correlation functions,
\beqa
\xi_\pm(\theta) 
&=&
\langle \gamma_t(0)\gamma_t(\boldsymbol{\theta})\rangle
\pm \langle \gamma_\times(0)\gamma_\times(\boldsymbol{\theta})\rangle, \label{eq:xi_pm_estimator} \\
&=&
\int \frac{\mathrm{d}\ell \ell}{2\pi} J_{0,4}(\ell \theta) \, P_\kappa(\ell),\label{eq:xi_pm_theory}
\eeqa
where the tangential ($t$) and cross ($\times$) components of the shear are defined relative to the direction connecting each pair of galaxies. In Eq.~(\ref{eq:xi_pm_theory}), the zeroth‑order Bessel function $J_0$ of the first kind applies to $\xi_+$, while the fourth‑order Bessel function $J_4$ applies to $\xi_-$. Because the estimators in Eq.~(\ref{eq:xi_pm_estimator}) depend only on the angular separation between galaxy pairs and are straightforward to measure observationally \cite{Schneider:2002jd}, these two‑point correlation functions form the foundation of most current weak‑lensing cosmological analyses.

\subsection{Mass mapping and beyond two-point statistics}\label{subsec:mass_map}

Reconstruction of the convergence (projected mass density) field from observed galaxy ellipticities was first proposed by Kaiser and Squires \cite{Kaiser:1992ps} and later applied to real data \cite{Fahlman:1994np}. However, it was subsequently shown that direct reconstruction techniques suffer from boundary artifacts when applied to surveys with finite sky coverage \cite{Seitz:1995dq}. This limitation makes pixel‑level reconstruction of convergence maps challenging in practice.

As an alternative, many weak‑lensing analyses focus on the \textit{smoothed} convergence field rather than on the raw reconstruction. This approach was introduced by Schneider \cite{Schneider:1996ug} and has since become standard, primarily because smoothing suppresses shape noise and mitigates boundary effects. 
The smoothed convergence field is defined as a filtered version of the convergence,
\beqa
{\cal K}(\boldsymbol{\theta})
= \int {\rm d}^{2}\phi \, \kappa(\boldsymbol{\theta}-\boldsymbol{\phi}) \, U(\boldsymbol{\phi}),
\label{eq:ksm_u}
\eeqa
where $U$ is a smoothing kernel. Equivalently, the same quantity can be obtained by filtering the tangential shear field \cite{Schneider:1996ug}. In practical applications, simple filter choices are commonly adopted, with the truncated Gaussian filter being one of the most widely used,
\beqa
U(\theta)
= \frac{1}{\pi\theta_G^{2}} \exp \left(-\frac{\theta^{2}}{\theta_G^{2}}\right)
\frac{1}{\pi\theta_o^{2}}
\left( 1 - \exp \left(-\frac{\theta_o^{2}}{\theta_G^{2}}\right)\right),
\label{eq:filter_gamma}
\eeqa
for $\theta\le\theta_o$ and zero otherwise. Typical choices such as 
$\theta_G=1\,\mathrm{arcmin}$ and $\theta_o=15\,\mathrm{arcmin}$ are optimized for detecting massive galaxy clusters in modern ground‑based weak‑lensing surveys \cite{Hamana:2003ts}.

Given smoothed convergence fields, one can extract a variety of summary statistics that probe cosmological information inaccessible to two-point correlation functions alone. Here we provide a concise overview of five representative statistics that have been extensively discussed in the literature.

\subsubsection*{Probability distribution function}
The one-point probability distribution function (PDF) of the convergence field captures essential information about the non-Gaussian nature of the underlying matter distribution \cite{Bernardeau:2000et, Liu:2018dsw, Barthelemy:2020yva, Thiele:2020rig, Boyle:2020bqn, Giblin:2022ucn}. Unlike two-point statistics, the convergence PDF incorporates contributions from moments of all orders, making it a sensitive probe of nonlinear structure formation on a given smoothing scale. Furthermore, as a purely local statistic, the PDF is straightforward to estimate from both simulations and observational data, even in the presence of irregular survey geometries \cite{DES:2017eav, DES:2017hhj, Thiele:2023gqr}.

\subsubsection*{Moments}
While the shear and convergence two-point correlation functions—or equivalently the power spectra—quantify only the variance of the lensing fields, higher-order moments offer access to additional non-Gaussian information. For a smoothed convergence field, moments may be written as
$\langle \kappa_n \rangle(\theta) = \int d\kappa \, \kappa^{n} \, {\rm Prob}(\kappa)$
assuming zero mean. In a Gaussian one-point distribution, 
the variance $\langle \kappa_2 \rangle = \sigma^2$ completely characterizes the field. 
Deviations from Gaussianity are conventionally described by the skewness 
$\langle \kappa_3\rangle / \sigma^3$ and the kurtosis $\langle \kappa_4\rangle / \sigma^4 - 3$. Although low-order cumulants provide a compact summary of the PDF \cite{Takada:2002ee, Jarvis:2003wq}, higher cumulants become increasingly sensitive to the distribution's tails, making them difficult to predict accurately in practice.
Nevertheless, recent studies that carefully control small‑scale physics have demonstrated that three‑point weak‑lensing statistics can break parameter degeneracies inherent in two‑point analyses \cite{DES:2021lsy, DES:2025llp}.

\subsubsection*{Peak counts}
Local maxima in convergence maps—known as peak counts—have emerged as a particularly effective summary statistic for capturing non-Gaussian information beyond the power spectrum \cite{Liu:2014fzc, Hamana:2015bwa, DES:2016jfa, Shan:2017mgz, Martinet:2017rqp, Harnois-Deraps:2020pvj, DES:2021epj, Liu:2022gnc, Marques:2023bnr}. High peaks (typically those height being $\gtrsim 4\sigma$) are associated with individual massive dark-matter halos \cite{Hamana:2003ts, Hennawi:2004ai, Maturi:2004rn, Marian:2011rg, Hamana:2012ur}, whereas peaks of more moderate heights reflect line-of-sight projections of large-scale structure or contributions from multiple halos \cite{Yang:2011zzn, Liu:2016xjb, Sabyr:2021vpr}. As a result, peak statistics serve as an intuitive and physically meaningful tool for studying nonlinear structure formation.

\subsubsection*{Minkowski functionals}
Minkowski functionals (MFs) provide a complementary means of characterizing the morphology of smoothed random fields. For two-dimensional convergence maps, the three MFs
—$V_0$, $V_1$, and $V_2$—correspond respectively to the area above a threshold 
$\kappa_{\mathrm{thre}}$, the total contour length, and the integrated geodesic curvature of the contours. 

Notably, $V_2$ effectively tracks the number of connected excursion sets 
and therefore becomes closely related to peak counts at high thresholds.
MFs can be computed perturbatively when the field displays weak non-Gaussianity \cite{Matsubara:2003yt, Munshi:2011wu, Matsubara:2020fet}, but analytic predictions remain challenging for strongly nonlinear regimes \cite{Petri:2013ffb}. Simulation-based approaches have demonstrated that MFs can extract cosmological information beyond that accessible to the power spectrum alone \cite{Kratochvil:2011eh, Shirasaki:2013zpa, Petri:2013ffb, Armijo:2024ujo}.

\subsubsection*{Scattering transform coefficients}
The scattering transform (ST) provides a mathematically well-controlled method for extracting information from high-dimensional fields. It possesses desirable properties such as translational invariance, non-expanding variance, and stability under small deformations \cite{Cheng:2021xdw}. 
The ST constructs a hierarchy of transformed fields through repeated wavelet convolutions and modulus operations; the expected values of these fields serve as informative summary statistics of the original lensing map.

Conceptually, the ST performs operations akin to those used in convolutional neural networks (CNNs), relying on localized filters and nonlinearities. Indeed, Ref.~\cite{Cheng:2020qbx} showed that ST-based summary statistics can deliver cosmological constraints comparable to those obtained with CNNs \cite{Ribli:2019wtw} for weak-lensing mass maps. Importantly, unlike CNNs—which require computationally costly training on many simulations—the ST demands only that its expected coefficients be modeled as a function of cosmology, similar to other summary statistics. The constraining power of the ST has been demonstrated with real weak-lensing data \cite{Cheng:2024kjv}.

\section{Common practice}
\subsection{Cosmological parameter inference under Gaussian likelihoods}
One of the central goals of weak gravitational lensing is the inference of cosmological parameters. In practice, the parameters of interest $q$ are inferred from the observational data vector $D_\mathrm{obs}$ through Bayes’ theorem,
\beqa
\mathrm{Prob}(q|D_\mathrm{obs}) \propto \mathrm{Prob}(D_\mathrm{obs}|q)\mathrm{Prob}(q),
\label{eq:bayes_theorem}
\eeqa
where $\mathrm{Prob}(q|D_\mathrm{obs})$ is the posterior distribution,
$\mathrm{Prob}(D_\mathrm{obs}|q)$ is the likelihood, and
$\mathrm{Prob}(q)$ denotes the prior encoding existing physical knowledge.
Here, $q$ generally includes not only the baseline cosmological parameters (e.g., $\Omega_\mathrm{m0}$ and $\sigma_8$) but also nuisance parameters describing modeling uncertainties—such as shear measurement biases, photometric-redshift errors, and baryonic effects on the matter power spectrum.

Inference under Eq.~(\ref{eq:bayes_theorem}) requires a forward model of the likelihood term $\mathrm{Prob}(D_\mathrm{obs}|q)$. A widely adopted assumption in weak-lensing analyses is that the likelihood can be approximated as multivariate Gaussian:
\beqa
\mathrm{Prob}(D_\mathrm{obs}|q)
&=& (2\pi)^{-N/2} \det(\Gamma)^{-1/2} \nonumber \\
&&
\times
\exp\left[-\frac{1}{2}\left(D_\mathrm{obs}-f(q)\right)^{T}\Gamma^{-1}\left(D_\mathrm{obs}-f(q)\right)\right],
\label{eq:Gaussian_likelihood}
\eeqa
where $N$ is the length of the data vector,
$\Gamma$ is the covariance matrix describing statistical uncertainties,
and $f(q)$ is the theoretical model prediction for the chosen summary statistic.

Equation~(\ref{eq:Gaussian_likelihood}) provides a useful starting point for cosmological inference, but practical challenges emerge in modeling both the mean vector $f(q)$ 
and the covariance $\Gamma$. Because weak-lensing observables are inherently nonlinear, neither quantity is straightforward to compute analytically. Below we summarize the current status of modeling $f$ and $\Gamma$ for various weak-lensing summary statistics.

\subsubsection*{Model functions}
For two-point statistics—most notably the convergence power spectrum $P_\kappa$—the modeling is well established. A formal decomposition of modeling uncertainties separates errors arising from the lensing kernel $W_\kappa$ and from the matter power spectrum $P_\mathrm{m}$. The former is dominated by uncertainties in the photometric redshift distribution of source galaxies, typically calibrated using external datasets and data-driven techniques \cite{Newman:2022rbn}. The latter relies on large suites of cosmological simulations. Foundational training datasets have been constructed from gravity-only N-body simulations \cite{Smith:2002dz, Heitmann:2008eq, Heitmann:2009cu, Lawrence:2009uk, Agarwal:2012ew, Takahashi:2012em, Agarwal:2013aea, Euclid:2018mlb, Angulo:2020vky, Moran:2022iwe, Chen:2025ugn}, while considerable effort has gone into modeling baryonic effects using hydrodynamical simulations and semi-analytic prescriptions \cite{Rudd:2007zx, vanDaalen:2011xb, Fedeli:2014gja, Osato:2015lja, Mead:2015yca, Mead:2016zqy, Chisari:2018prw, Barreira:2019ckp, Chisari:2019tus, Arico:2019ykw, Mead:2020vgs, Osato:2020sxo, Giri:2021qin, Acuto:2021yjm, Salcido:2023etz, Schaller:2024jiq, Schneider:2025zca, Kovac:2025zqy}.

For summary statistics beyond the two-point level, simulation-based emulators are commonly employed. Ensemble averages over mock surveys allow the prediction of arbitrary statistics, provided that the simulations reproduce realistic survey conditions. These include varying survey depth, irregular masks, source-selection effects \cite{Hamana:2001kd, Schmidt:2009rh, Schmidt:2009ri, Krause:2009yr, Schmidt:2010ex, Liu:2013yna, Deshpande:2019sdl, Duncan:2021jxl}, inhomogeneous source distributions \cite{Hamana:2000wb, Schneider:2001af, Valageas:2013qfa, Yu:2014iea, DES:2023ycm, KiDS-1000:2024rdr, Duncan:2024dzj}, and intrinsic galaxy alignments \cite{Troxel:2014dba, Chisari:2025gsy}.
At present, fully realistic mock datasets incorporating all these effects simultaneously are not yet available, raising concerns about the robustness of cosmological inference using higher-order statistics.

\subsubsection*{Covariance matrices}
A simple approach to estimating covariance matrices is based on resampling the observational data. Rotating individual galaxy shapes erases cosmological shear while preserving measurement noise, allowing direct estimation of shape-noise covariance. The delete-one jackknife method is also widely used for two-point statistics \cite{Friedrich:2015nga, Shirasaki:2016fuf}. In this method, the survey area is divided into subregions and the statistic is recomputed after removing each region. Jackknife covariances capture finite-sample effects absent in shape-noise methods but become unreliable when the statistic probes scales larger than the jackknife subregion, causing the loss of large-scale cosmological information.

A more direct approach is to estimate covariance matrices from many numerical simulations. If sufficiently many realizations of $D_\mathrm{obs}$ are available, 
the covariance can be estimated via the standard sample-covariance formula. 
However, stable estimation requires roughly $10–100\, N$ realizations for a data vector of size $N$ \cite{Hartlap:2006kj, Taylor:2012kz, Dodelson:2013uaa}, making this approach computationally prohibitive with current large-scale simulations \cite{Takahashi:2017hjr, Shirasaki:2019gya}.

Analytic covariance models calibrated against simulations exist for two-point shear statistics \cite{Sato:2009ct, Pielorz:2009sv, Takada:2013wfa, Krause:2016jvl, Mohammed:2016sre, Barreira:2017fjz}, but extending such models to higher-order statistics remains under development \cite{Kayo:2012nm, Shirasaki:2015dga, Uhlemann:2022znd, Bayer:2022nws, Linke:2022xnl, Wielders:2025ust}.
To alleviate computational demands, several strategies have been explored, including data compression techniques \cite{Heavens:2017efz, Bellini:2019lnn, Heavens:2020spq, Philcox:2020zyp, Ferreira:2020bgt, Heavens:2023liw, Sugiyama:2025bhl}, recycling limited simulation samples \cite{Schneider:2011wf, Morrison:2013tqa, Petri:2016wlu}, and hybrid methods that combine approximate analytic covariances with simulation-based estimates \cite{Friedrich:2017xxx, Hall:2018umb, Looijmans:2024sna}.

\subsection{Generative models of weak lensing convergence}\label{subsec:gen_model_WL}

A generative model for weak-lensing mass maps plays an essential role in modern weak-lensing analyses. In practice, such models are in high demand for several key reasons:

\begin{itemize}
    \item For many non-Gaussian summary statistics proposed in the literature, reliable analytic predictions for their cosmological dependence are still lacking. As a result, ensemble averages computed from a large set of generated weak-lensing convergence maps remain the standard approach for studying their sensitivity to cosmological parameters.
    \item  Precise error estimation is crucial for robust cosmological inference. The strongly non‑Gaussian nature of weak-lensing mass maps, combined with several observational systematics, renders analytical error predictions unreliable. The best available strategy is to perform suites of numerical simulations incorporating all relevant physical processes and then generate realistic mock lensing data, including observational effects such as survey masks, shape noise, and PSF residuals \cite{LSSTDarkEnergyScience:2019hkz, Shirasaki:2019gya}. 
    \item Given recent discussions about potential cosmological tensions (see Ref.~\cite{Abdalla:2022yfr}), blind analyses have become standard practice to minimize human‑induced biases. Current blinding schemes typically modify observed galaxy shapes using a single scaling parameter, effectively altering the amplitude of the lensing power spectrum \cite{KiDS:2020suj,DES:2021vln,Dalal:2023olq,Li:2023tui}. More sophisticated blinding procedures are needed to test the accuracy of theoretical templates for various lensing statistics. A ``blinded challenge" based on analyzing generated weak-lensing maps with hidden input cosmological parameters provides one promising direction (see Ref.~\cite{Nishimichi:2020tvu} for an example in galaxy clustering). Rapid and accurate generative models of weak-lensing mass maps would make such challenges significantly more practical.
\end{itemize}

Ray‑tracing simulations remain the most robust generative models available. They compute lensing by tracing light rays through matter fields extracted from cosmological simulations. However, these simulations require substantial computational resources because each realization depends on full N‑body or hydrodynamical simulations. As a consequence, only of order $10^2$–$10^4$ realizations can typically be produced per year \cite{Sato:2009ct,Dietrich:2009jq,Harnois-Deraps:2012ixt,Liu:2017now,Takahashi:2017hjr,Harnois-Deraps:2018bcv,Harnois-Deraps:2019rsd,Shirasaki:2019wxk,Osato:2020sxo,Kacprzak:2022pww,Ferlito:2023gum, DES:2024xij}.

To reduce this computational burden, several alternative approaches have been proposed. These include multivariate log-normal models \cite{Taruya:2002vy, Xavier:2016elr, Makiya:2020iai} and their extensions \cite{Yu:2016qoq, Shirasaki:2016vve, Tessore:2023zyk, Zhong:2024wdk}, approximate gravity solvers \cite{Izard:2017kma, Sgier:2018soj, Bohm:2020ilt}, and halo‑based prescriptions \cite{Giocoli:2017nnq, Giocoli:2020pfb}. While these methods drastically reduce runtime, their predictive accuracy for highly non‑Gaussian observables is generally limited.

A representative example of a fast generative model is the log-normal model, which assumes that the real-space convergence field follows the one-point probability distribution
\beqa
\mathrm{Prob}(\kappa) &=& \frac{1}{\sqrt{2\pi} \sigma_{\ln}}\frac{1}{\kappa+|\kappa_\mathrm{min}|} \nonumber \\
&& \times \exp\left\{-\frac{\left[\ln(1+\kappa/|\kappa_\mathrm{min}|)+\sigma^2_{\ln}/2\right]^2}{2\sigma^2_{\ln}}\right\},
\label{eq:lognormal_pdf}
\eeqa
for $\kappa > \kappa_\mathrm{min}$, and zero otherwise. 
Here the variance parameter $\sigma^2_{\ln}$ is defined as
\beqa
\sigma^2_{\ln} \equiv \ln \left(1+\frac{\langle \kappa^2 \rangle}{|\kappa_\mathrm{min}|^2}\right),
\eeqa
where $\langle \kappa^2 \rangle$ is determined by the convergence power spectrum $P_\kappa$.

This model may be reformulated by introducing a Gaussian auxiliary field $y$, 
defined through the transformation
\beqa
y(\boldsymbol{\theta}) = \ln\left(1+\frac{\kappa(\boldsymbol{\theta})}{|\kappa_\mathrm{min}|}\right)+\frac{\sigma^2_{\ln}}{2}.
\label{eq:lognormal2Gaussian}
\eeqa
By construction, $y$ has zero mean and unit variance. 
Using Eq.~(\ref{eq:lognormal2Gaussian}), the two-point correlation of $y$ 
is related to that of $\kappa$ via
\beqa
\xi_{y}(\theta) &=& \langle y(\boldsymbol{\theta}+\boldsymbol{\phi}) y(\boldsymbol{\phi}) \rangle, \\
\xi_{\kappa}(\theta) &=& \langle \kappa(\boldsymbol{\theta}+\boldsymbol{\phi}) \kappa(\boldsymbol{\phi}) \rangle \nonumber \\ 
&=& \frac{1}{2\pi[1-\xi^2_y(\theta)]^{1/2}}
\int_{-\infty}^{\infty}\mathrm{d}u_1 \int_{-\infty}^{\infty}\mathrm{d}u_2 \,
e^{-u_1^2/[2(1-\xi^2_y)]}\, e^{-u_2^2/2} \nonumber \\
&&
\qquad \qquad \qquad
\qquad \qquad \qquad
\times {\cal F}(u_1+\xi_y u_2){\cal F}(u_2),
\label{eq:kappa_2pcf_lognormal}
\eeqa
where ${\cal F}$ is the inverse of the transformation in Eq.~(\ref{eq:lognormal2Gaussian}):
${\cal F}(u)=|\kappa_\mathrm{min}| e^{u\sigma_{\ln}-\sigma_{\ln}^2/2}-|\kappa_\mathrm{min}|$.
The correlation function $\xi_{\kappa}(\theta)$ 
can also be obtained from the convergence power spectrum $P_\kappa$ via
\beqa
\xi_{\kappa}(\theta) = \int\frac{\mathrm{d}^2\ell}{(2\pi)^2}\, P_\kappa(|\boldsymbol{\ell}|) e^{-i\boldsymbol{\ell}\cdot\boldsymbol{\theta}}.
\eeqa
Thus Eq.~(\ref{eq:kappa_2pcf_lognormal}) determines $\xi_{y}(\theta)$ 
uniquely for a given $P_\kappa$. In this way, the log-normal generative model is fully specified by only two ingredients: the minimum convergence $\kappa_\mathrm{min}$ and the convergence power spectrum $P_\kappa$.


\begin{figure}[t]
\includegraphics[width=0.8\linewidth, bb=200 0 1440 1200]{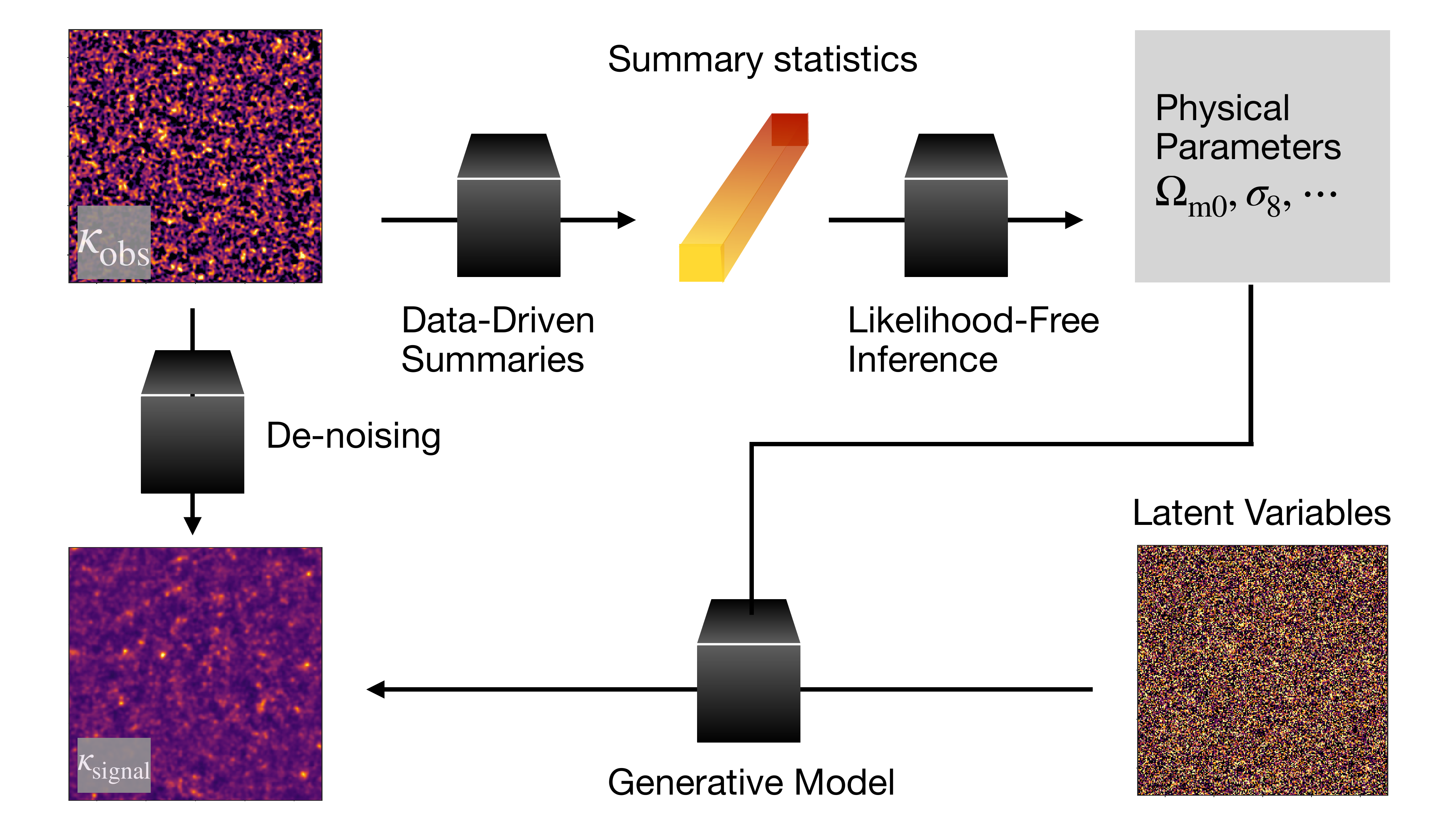}
%
%
\caption{
A schematic summary of machine‑learning applications to weak‑lensing cosmology discussed in this chapter.
The top‑left panel shows a noisy convergence field, $\kappa_\mathrm{obs}$, while the bottom‑left panel displays its noiseless counterpart. To infer cosmological parameters of interest (e.g., $\Omega_\mathrm{m0}$ and $\sigma_8$, illustrated in the top‑right panel), it is essential to extract compact yet informative features from the observable $\kappa_\mathrm{obs}$. This feature extraction can be performed in a data‑driven manner using neural networks, as discussed in the first half of Section~\ref{subsec:feature_extraction}.
A practical cosmological inference further requires information about likelihood functions. Neural networks can efficiently estimate credible regions of parameters by directly evaluating the posterior distribution given an observation, an approach known as likelihood‑free inference, introduced in the second half of Section~\ref{subsec:feature_extraction}. Another key application is noise reduction (de‑noising) of $\kappa_\mathrm{obs}$ to clarify the spatial information of large‑scale structures; recent advances in this area are summarized in Section~\ref{subsec:denoising}. Finally, Section~\ref{sebsec:deep_gen_model_WL} provides a comprehensive review of neural‑network-based generative models that produce noiseless convergence fields from random latent variables.
}
\label{fig:1}       
\end{figure}

\section{Applications of machine learning}

Weak gravitational lensing directly probes the projected mass distribution of the Universe, but extracting cosmological information from noisy data is challenging. This section reviews how machine‑learning methods enhance weak‑lensing analyses through data‑driven inference, noise reduction, and generative modeling, enabling more efficient use of lensing datasets (Figure~\ref{fig:1}).

\subsection{Feature extractions for cosmological parameter inference}\label{subsec:feature_extraction}

\subsubsection*{Data-driven summaries}
Machine learning provides a flexible framework for extracting physical information from data without imposing strong prior assumptions or handcrafted summary statistics. Motivated by rapid developments in deep learning, a variety of data-driven summary statistics have been proposed for cosmological inference problems.

A common setup is to train a model that maps observational inputs directly to cosmological parameters such as $\Omega_\mathrm{m0}$ and $\sigma_8$. A pioneering study by Ref.~\cite{Gupta:2018eev} demonstrated that a two-dimensional convolutional neural network (CNN) can efficiently compress simulated noiseless weak-lensing maps into constraints on $(\Omega_\mathrm{m0}, \sigma_8)$, achieving an improvement by a factor of $\sim 5$ over traditional two-point statistics. Although the inclusion of shape noise reduces the gains from CNNs, Ref.~\cite{Fluri:2018hoy} showed that CNNs can still outperform standard power-spectrum analyses when an appropriate smoothing scale is adopted for constructing convergence maps. The magnitude of this improvement depends sensitively on the observational noise level. In particular, Ref.~\cite{Ribli:2019wtw} emphasized that CNNs are especially promising for future surveys with high source galaxy number densities.

Applications to real data have further tested the robustness of CNN-based approaches. Studies using catalogs from the Kilo Degree Survey \cite{Fluri:2019qtp}, the Subaru Hyper Suprime-Cam Survey \cite{Lu:2023dep}, and the Dark Energy Survey \cite{DES:2024xij} have shown that CNNs can provide tighter cosmological constraints than the standard power spectrum, even in the presence of realistic systematics.

Detailed investigations in Ref.~\cite{Ribli:2018kwb} revealed that CNNs may capitalize on subtle features—such as gradients around convergence peaks—that are typically missed by traditional summary statistics. Using saliency-based analyses, Ref.~\cite{ZorrillaMatilla:2020doz} found that CNNs predominantly extract information from high-convergence regions when shape noise is present. To further enhance performance, Ref.~\cite{Zhong:2024qpf} introduced random permutation blocks, which randomize pixel locations within intermediate CNN feature maps. This procedure regularizes the network in low–signal-to-noise regimes and effectively augments the training data. The choice of loss function also plays an important role: Ref.~\cite{Sharma:2024pth} demonstrated that alternative training objectives can lead to improved constraints, pointing toward the need for physics-motivated training strategies.

Most recently, Ref.~\cite{Makinen:2024bff} proposed an effective method for combining CNN outputs with traditional power-spectrum information in a non-overlapping way. This hybrid approach clarifies the physical origin of the information captured by CNNs and offers a pathway to systematically integrate machine-learning summaries into standard cosmological analyses.

\begin{figure}[t]
\includegraphics[width=0.8\linewidth, bb=200 0 1440 1200]{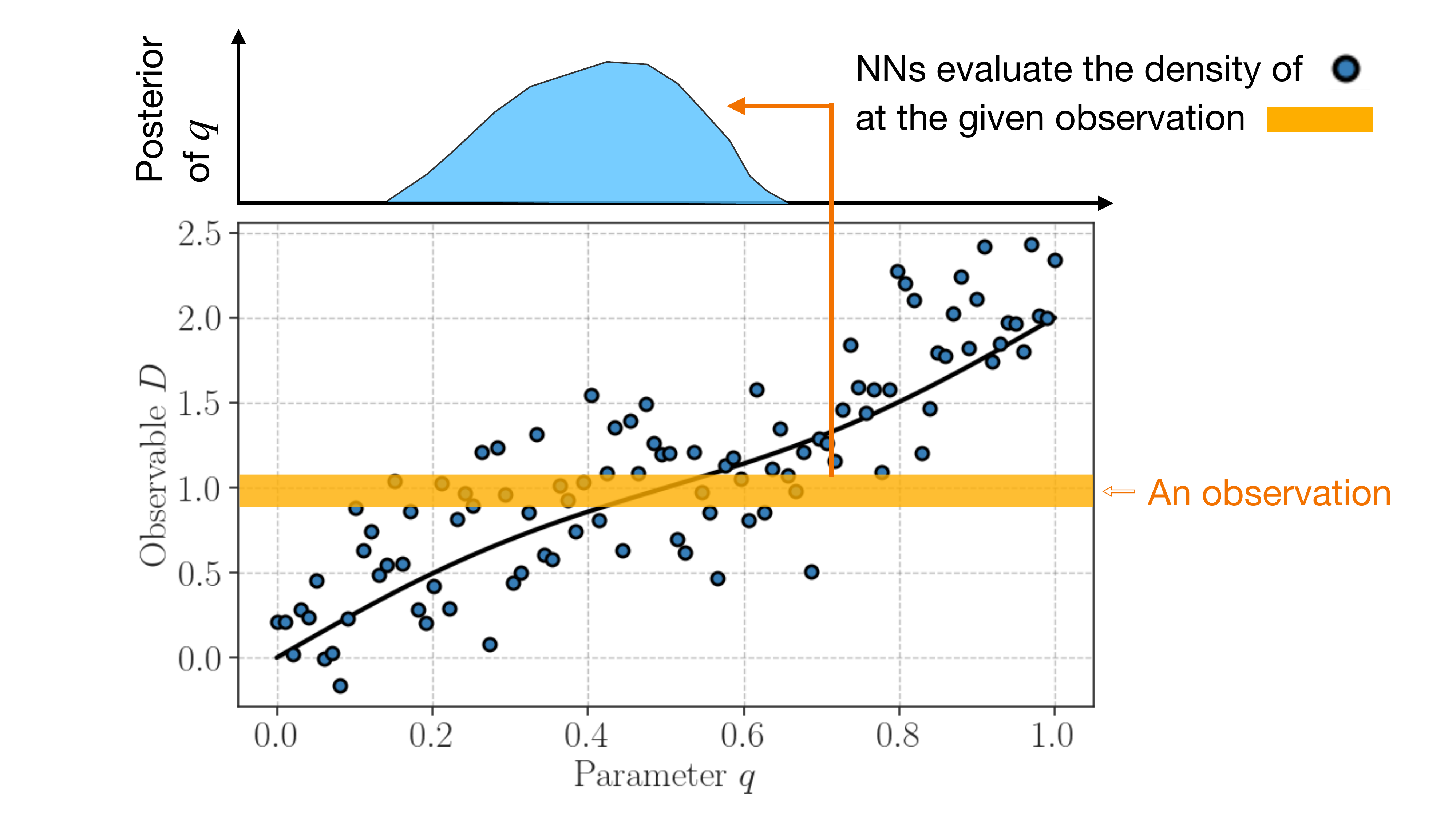}
%
%
\caption{
Conceptual illustration of likelihood‑free inference in a one‑dimensional example.
The goal is to infer a parameter $q$ from an observed data value $D$. Assuming that the data $D$ can be forward‑modeled with reasonable accuracy for any given $q$, the blue points represent samples generated by the simulator. Likelihood‑free inference then aims to estimate the posterior density of $q$ conditioned on the observed value $D = D_\mathrm{obs}$, highlighted in orange.
}
\label{fig:2}       
\end{figure}

\subsubsection*{Likelihood-free inference}
Another important machine-learning application to cosmological inference is the development of likelihood-free inference (LFI). Unlike traditional analyses that assume a Gaussian likelihood—such as the form in Eq.~(\ref{eq:Gaussian_likelihood})—LFI aims to estimate the posterior distribution of cosmological parameters directly from data, without requiring an explicit likelihood model. Conventional Gaussian-likelihood inference requires building models for the mean and covariance of a chosen summary statistic, typically using a combination of semi-analytic calculations and simulation-based calibration. LFI circumvents this requirement entirely.

The key idea behind LFI is that one can forward-simulate mock data vectors $D_\mathrm{obs}$ at specified parameter values $q$. If the simulation process approximates the true likelihood $\mathrm{Prob}(D_\mathrm{obs}|q)$ sufficiently well, then parameter inference reduces to a non‑parametric density estimation problem. This approach is especially valuable when summary statistics lack closed-form predictions, or when the Gaussian-likelihood assumption is questionable. LFI also avoids the challenges of sampling from high-dimensional Bayesian hierarchical models, which can be computationally prohibitive. However, one trade-off is that classical goodness‑of‑fit diagnostics—such as $\chi^2$ tests—are not naturally defined within the LFI framework. Figure~\ref{fig:2} illustrates the basic concept of LFI using a simple one-dimensional example.

Neural network–based likelihood estimators, often called neural density estimators, have been applied to a variety of weak-lensing survey simulators, sometimes in combination with compressed summary statistics \cite{Jeffrey:2020xve, Lin:2022ayr, DES:2024xij, vonWietersheim-Kramsta:2024cks, DES:2024jgw, Novaes:2024dyh}. The performance and robustness of LFI have been further assessed using controlled non‑Gaussian likelihoods \cite{Homer:2024cwg}, showing that the credible intervals obtained via LFI may become wide when the number of simulation draws is insufficient. Strategies such as pre‑training with a large number of inexpensive (``cheap") simulations followed by fine‑tuning with a smaller set of accurate simulations have been proposed to accelerate convergence \cite{Saoulis:2025bug}. 

Ref.~\cite{Diao:2025szg} demonstrated that a neural density estimator can be repurposed for out-of-distribution (OoD) detection, thereby providing a practical means of assessing the validity domain of the simulator itself. 
The idea is to compute the posterior predictive distribution, defined as
\beqa
\mathrm{Prob}(D_\mathrm{sim}|D_\mathrm{obs}) =
\int \mathrm{d}q \, \mathrm{Prob}(D_\mathrm{sim}|q)\,\mathrm{Prob}(q|D_\mathrm{obs}),
\eeqa
where the posterior $\mathrm{Prob}(q|D_\mathrm{obs})$ is obtained via LFI, 
and $\mathrm{Prob}(D_\mathrm{sim}|q)$ corresponds to the forward-model simulator evaluated at parameters $q$. 
If the cumulative posterior predictive probability,
\beqa
t(D_\mathrm{obs}) = \int_{D_\mathrm{sim}>D_\mathrm{obs}} \mathrm{d}D_\mathrm{sim}\, \mathrm{Prob}(D_\mathrm{sim}|D_\mathrm{obs}),
\eeqa
approaches either 0 or 1, the observed data vector $D_\mathrm{obs}$ is identified as an OoD sample. This indicates that the simulator is unable to reproduce the observed data under any reasonable parameter choice, providing an effective diagnostic for simulator validity.

\subsection{Denoising lensing mass maps with convolutional neural networks}\label{subsec:denoising}

Mapping the cosmic mass density with weak-lensing measurements provides a unique way to visualize the spatial distribution of invisible dark matter. In practice, however, this approach is severely affected by observational noise. The dominant difficulty arises from the fact that the original, unlensed images of individual galaxies are never observable. As a consequence, the cosmic-shear estimator (Eq.~[\ref{eq:eps_obs_g_eps_int}]) is inevitably contaminated by intrinsic galaxy shapes, commonly referred to as shape noise. Moreover, the typical amplitude of intrinsic ellipticities is much larger than the cosmic-shear signal of interest. As a result, conventional mass-mapping analyses (Section~\ref{subsec:mass_map}) are effective for identifying highly massive structures such as galaxy clusters on an individual basis, but they fail to recover the diffuse mass components that dominate most pixels in a lensing convergence map.

To overcome these limitations, deep-learning approaches have attracted considerable attention. Ref.~\cite{Shirasaki:2018thk} first proposed reducing shape noise in lensing mass maps using 
deep-learning-based image-to-image transformations. The key idea is that an observed convergence map $\kappa_\mathrm{obs}$ can be expressed as the sum of a signal and noise contribution,
\beqa
\kappa_\mathrm{obs} = \kappa_\mathrm{signal} + \kappa_\mathrm{noise},
\eeqa
where $\kappa_\mathrm{signal}$ arises from gravitational lensing by large-scale structure, while $\kappa_\mathrm{noise}$ reflects intrinsic galaxy shapes. This decomposition motivates the objective of estimating $\kappa_\mathrm{signal}$ directly from a noisy input map $\kappa_\mathrm{obs}$. Because the noise-injection process can be forward-modeled (see Eq.~[\ref{eq:eps_gamma_in_WL}]), modern deep neural networks can be trained on paired datasets $(\kappa_\mathrm{obs}, \kappa_\mathrm{signal})$ to learn the conditional probability distribution $\mathrm{Prob}(\kappa_\mathrm{signal}|\kappa_\mathrm{obs})$.

\begin{figure}[t]
\includegraphics[width=0.8\linewidth, bb=240 100 1440 800]{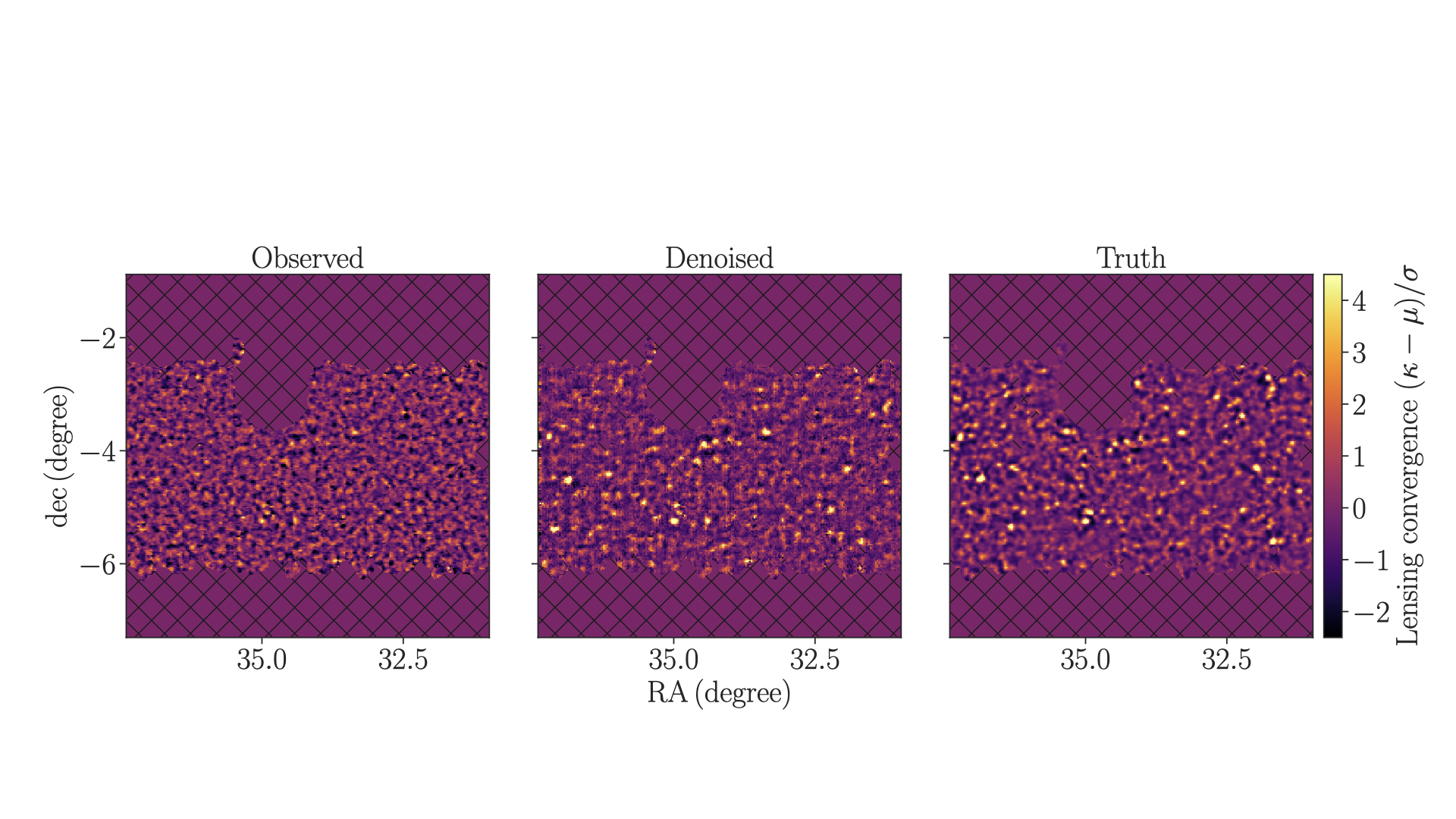}
%
%
\caption{
An example of noise reduction applied to lensing convergence maps obtained from the Hyper Suprime-Cam (HSC) imaging survey \cite{Shirasaki:2019wxk}.
A deep‑learning network is used to estimate the noise component present in the observed data and to remove it from the observed convergence maps. The figure shows a validation using test data that were not used during the training of the network. The left panel displays the observed convergence field including noise, while the right panel shows the noiseless convergence field arising purely from gravitational lensing. The middle panel presents the denoised convergence field produced by the neural network. In all panels, the convergence fields are normalized to have zero mean and unit standard deviation. Hatched regions indicate areas with no observed galaxies. Whereas the noisy map contains physically meaningful information only in pixels with very large convergence values, the denoised map successfully recovers the underlying large‑scale structure pattern across most of the field.
}
\label{fig:3}       
\end{figure}

In Ref.~\cite{Shirasaki:2018thk}, the authors further simplified the learning objective by training networks to produce a single plausible realization of $\kappa_\mathrm{signal}$ given $\kappa_\mathrm{obs}$. This formulation closely resembles classical image denoising problems in computer vision and naturally lends itself to conditional generative adversarial networks (cGANs). A related approach was adopted in Ref.~\cite{Jeffrey:2019fag}, where a CNN was used to reconstruct $\kappa_\mathrm{signal}$ from Dark Energy Survey science verification data. Ref.~\cite{Shirasaki:2019wxk} performed extensive stress tests using Subaru Hyper Suprime-Cam Survey data, demonstrating that cGAN-based denoising can recover small galaxy clusters that are otherwise hidden by shape noise.
Figure~\ref{fig:3} shows an example of the noise reduction of representative mock data 
for the Subaru Hyper Suprim-Cam Survey with a complex survey boundary. 

Ref.~\cite{Remy:2022ixn} argued that recently developed score-based generative models (also known as diffusion models) are better suited for learning $\mathrm{Prob}(\kappa_\mathrm{signal}|\kappa_\mathrm{obs})$ in a principled manner. Their carefully designed network architectures preserve Gaussian statistics on large cosmological scales, enabling efficient training even with a limited number of realizations. An updated architecture presented in Ref.~\cite{Whitney:2024pmf} introduced the capability to rapidly generate multiple realizations of $\kappa_\mathrm{signal}$ from a single noisy input using cGANs, allowing reconstruction uncertainties to be quantified. Ref.~\cite{Aoyama:2025aut} compared cGANs and diffusion models within a unified framework, highlighting several advantages of diffusion models, including improved training stability and direct sampling of the posterior distribution $\mathrm{Prob}(\kappa_\mathrm{signal}|\kappa_\mathrm{obs})$.

Finally, masked regions in real weak-lensing data present an additional challenge for deep-learning–based reconstruction when conventional CNNs are used. Furthermore, standard CNNs designed for flat two-dimensional images become inadequate for surveys covering a large portion of the sky. Ref.~\cite{Wang:2026ygy} demonstrated that transformer-based architectures can naturally accommodate survey masks and curved-sky effects, although their initial experiments were limited to training data based on simple, idealized Gaussian convergence fields.

\subsection{Deep generative models of weak lensing mass maps}\label{sebsec:deep_gen_model_WL}

\begin{figure}[t]
\includegraphics[width=0.6\linewidth, bb=200 200 1100 1100]{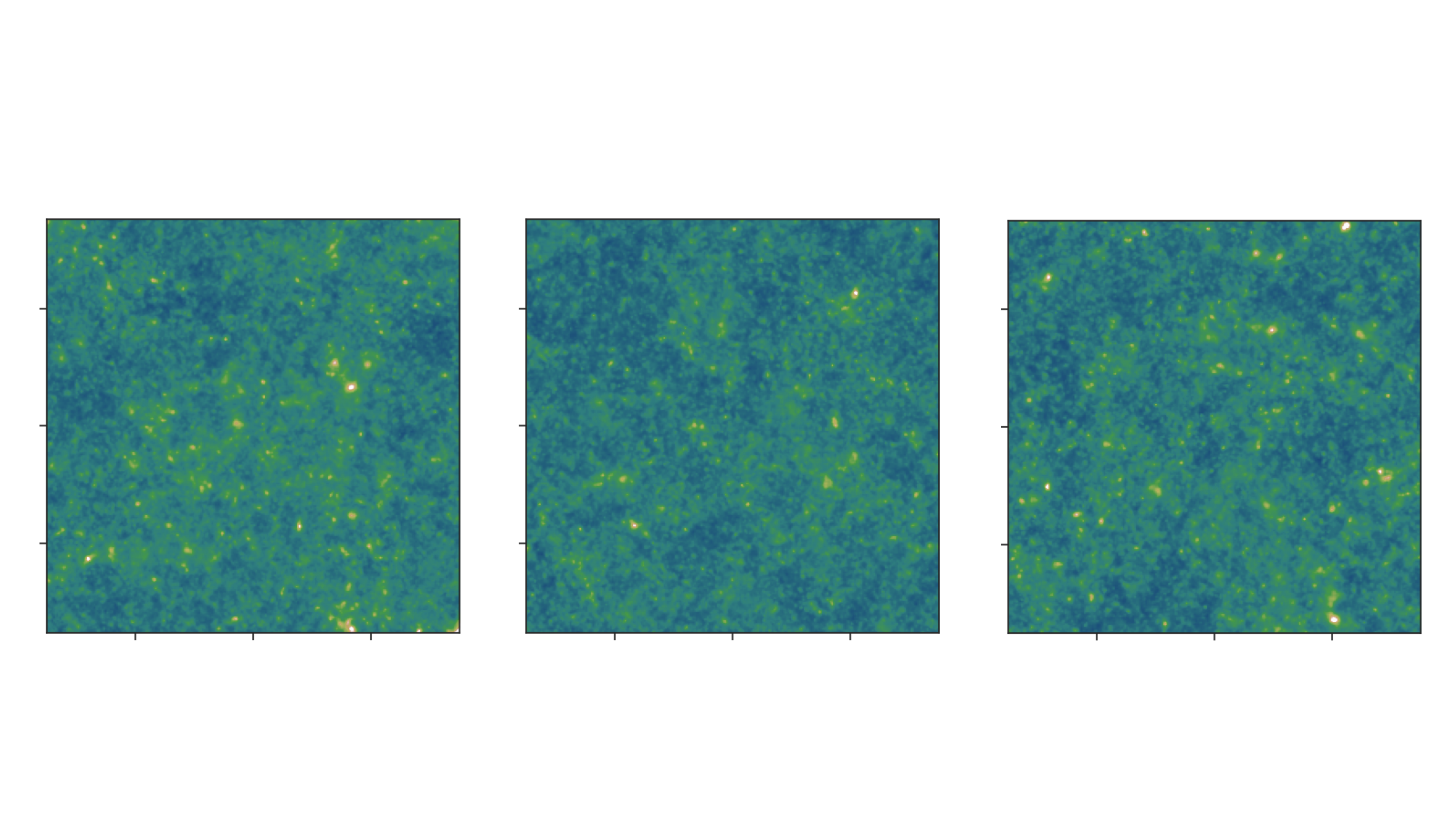}
%
%
\caption{
Examples of lensing convergence fields generated by neural networks \cite{Shirasaki:2023nnk}. Among the three images shown here, two are synthetic convergence maps produced by a neural‑network–based generative model, while one is a ``real'' map drawn from cosmological numerical simulations. We invite the reader to see whether they can tell which is which.
}
\label{fig:4}       
\end{figure}

Modern weak-lensing surveys require accurate and flexible generative models of weak-lensing observables, as discussed in Section~\ref{subsec:gen_model_WL}. However, many of the models proposed so far remain relatively inflexible, since they often rely on simple parametric forms or physically motivated approximations to generate observables. Deep-learning-based generative models offer a compelling alternative, as neural networks can act as universal function approximators, provided sufficiently large and representative training datasets are available (see Figure~\ref{fig:4} for a demonstration).

Ref.~\cite{Mustafa:2017bqp} first introduced a neural-network-based generative model for weak-lensing convergence fields, demonstrating that a generative adversarial network (GAN) can rapidly produce realistic lensing maps after appropriate training. This pioneering work motivated subsequent studies that extended GAN-based models by explicitly conditioning them on cosmological parameters \cite{Perraudin:2020gig}. Ref.~\cite{Tamosiunas:2020rvw} further examined interpolation in the latent space of GANs, showing that smooth transitions in the latent variables can generate plausible convergence fields corresponding to intermediate cosmological parameters that were not included in the training set.

In standard GAN implementations, latent variables are typically drawn from simple Gaussian noise distributions. Ref.~\cite{Shirasaki:2023nnk}, however, proposed an unpaired image-to-image translation approach using cycle-consistent GANs, enabling generative modeling with more physically interpretable latent variables. This formulation makes the input domain visually meaningful and allows the trained model to generate convergence maps with sky coverage larger than that of the training data, without requiring additional retraining. Ref.~\cite{Boruah:2024rgr} also demonstrated that carefully designed network architectures can significantly reduce the number of trainable parameters, showing that as few as $\mathcal{O}(10^3)$ parameters can suffice to generate realistic weak-lensing fields, compared to the $\mathcal{O}(10^7)$ parameters typically used in conventional CNN-based models.

While GANs can generate convergence maps very efficiently, their training is often unstable, and the resulting models may suffer from limited output diversity. These issues can be mitigated by diffusion models, which are grounded in a well-defined probabilistic framework and offer improved training stability while maintaining diversity in generated samples. Refs.~\cite{Boruah:2025zkn, Boruah:2025iwf} showed that a single diffusion model can be used both to generate realistic convergence maps and to denoise practical weak-lensing mass maps. Diffusion models are particularly competitive in producing high-resolution outputs, although their computational cost during generation is generally higher than that of GAN-based approaches.

Normalizing flows provide another promising framework for constructing generative models of cosmological fields, with the distinctive advantage of enabling exact evaluation of the data-generating probability density. These models assume an invertible mapping between latent Gaussian variables and the data space. This constraint limits the architectural flexibility of the neural networks used in the flow and necessitates careful tuning of network structures and training hyperparameters (e.g., optimizer learning rates). Nevertheless, Ref.~\cite{Dai:2022dso} demonstrated that conditional normalizing flows with symmetric architectures can successfully generate two-dimensional cosmic mass density fields without relying on complex neural networks. Ref.~\cite{Armijo:2026ucb} further noted that normalizing-flow-based models may struggle to reproduce correlated scatter among different weak-lensing summary statistics, highlighting the need for further investigations to fully capture the diversity of weak-lensing fields within this framework.

\section{Outlook}
In closing this chapter, we outline several promising directions for future applications of machine-learning methods in weak-lensing analyses. These include the development of field-level inference, the construction of lightweight and user-friendly generative models, and the reconstruction of three-dimensional large-scale structures using weak-lensing observables. Naturally, the perspectives presented here reflect the author’s interests; nevertheless, unforeseen and novel applications are very much anticipated.

\subsection*{Field level inferences}
Field-level inference—direct estimation of cosmological parameters from weak-lensing data without intermediate data compression—represents one of the ultimate goals in weak-lensing cosmology. By construction, this approach is lossless in terms of information extraction and therefore promises optimal cosmological constraints. Weak-lensing data are particularly well suited to this framework because the likelihood can be well defined: when intrinsic shape noise is approximated as a Gaussian random field in real space, masked regions and complex survey boundaries can be incorporated in a straightforward manner, making field-level inference practically feasible.

Initial studies of field-level inference in weak-lensing cosmology have relied on simplified models, such as log-normal approximations \cite{Fiedorowicz:2021clg, Boruah:2022lsu, Zhou:2023ezg, Boruah:2024tqp} or physically motivated perturbation-theory approaches \cite{Porqueres:2023drp}. While these models provide useful proof-of-concept demonstrations, their ability to extract information from small-scale nonlinear structures remains uncertain. A natural avenue for improvement is the extensive use of machine-learning generative models, which may enable the extraction of rich information from highly nonlinear structures such as galaxy clusters. However, such approaches require careful validation to ensure robustness and interpretability.

Another important extension of field-level inference lies in its joint application to weak lensing and complementary cosmological observables, such as galaxy positions. To date, no fully viable framework has been established for multi-field-level inference capable of simultaneously probing multiple tracers. Such an extension would represent a natural generalization of combined two-point analyses and could enable lossless, optimal inference in future galaxy surveys.

\subsection*{Lightweight neural networks for robust and efficient generative models}
Deep-learning generative models are often regarded as black-box tools. While their performance is highly attractive, limited interpretability has so far hindered their widespread adoption in precision cosmology. It is worth emphasizing, however, that cosmological observables are structurally simpler than many targets encountered in computer vision. The large-scale structure of the Universe is governed by physical symmetries—most notably statistical homogeneity and isotropy—which can, in principle, be exploited to construct interpretable yet flexible neural architectures.

A representative example is provided in Ref.~\cite{Dai:2022dso}, which demonstrated that incorporating the cosmological principle can substantially simplify model architectures. Extending such ideas to weak-lensing observables is particularly intriguing, since gravitational lensing probes highly nonlinear structure formation processes. A fundamental question in developing lightweight neural networks concerns the manifold hypothesis: can cosmological observables be effectively described using a small number of latent variables? If so, which physical processes determine or degrade the information encoded in these latent representations? Recent progress in deep-learning generative models, as summarized in Section~\ref{sebsec:deep_gen_model_WL}, suggests that weak-lensing convergence fields may indeed admit compact latent representations. However, the physical origin of this apparent efficiency remains poorly understood and merits further investigation.

\subsection*{Three-dimensional reconstructions of large-scale structures}

Most lensing mass-mapping analyses focus on reconstructing a single convergence field, corresponding to a two-dimensional projection of the cosmic mass distribution. Extending this framework to three-dimensional reconstructions is a natural and scientifically compelling goal. Early attempts at three-dimensional mass mapping using weak-lensing data employed Wiener filtering techniques \cite{Massey:2007wb, Oguri:2017vrv}, which provide stable reconstructions but tend to suppress small-scale information. Methods based on sparse priors \cite{Leonard:2013hia, Leonard:2015jha, Li:2021vsa} have enabled more detailed reconstructions in localized regions around galaxy clusters, although their practical field of view remains limited.

Deep-learning-based approaches offer a promising alternative, owing to their flexibility and computational efficiency. Nevertheless, careful formulation of learning objectives is essential to avoid overfitting, particularly given the limited size of realistic training datasets. Incorporating complementary information from galaxy distributions and other tracers of large-scale structure may help stabilize the training of deep-learning-based deprojection models and improve the reliability of three-dimensional reconstructions.

\begin{acknowledgement}
MS is supported by JSPS KAKENHI Grant Numbers JP24H00221 and JP24H00215.
MS also aknowledges research supports by JST BOOST, Japan Grant Number JPMJBY24D8.
\end{acknowledgement}







\begin{thebibliography}{99}








\bibitem{Seitz:1996vf}
C.~Seitz and P.~Schneider,
Astron. Astrophys. \textbf{318}, 687 (1997)
[arXiv:astro-ph/9601079 [astro-ph]].

\bibitem{Li:2021mvq}
X.~Li, H.~Miyatake, W.~Luo, S.~More, M.~Oguri, T.~Hamana, R.~Mandelbaum, M.~Shirasaki, M.~Takada and R.~Armstrong, \textit{et al.}
Publ. Astron. Soc. Jap. \textbf{74}, no.2, 421-459-459 (2022)
doi:10.1093/pasj/psac006
[arXiv:2107.00136 [astro-ph.CO]].

\bibitem{Limber:1954zz}
D.~N.~Limber,
Astrophys. J. \textbf{119}, 655 (1954)
doi:10.1086/145870


\bibitem{Schneider:2002jd}
P.~Schneider, L.~van Waerbeke, M.~Kilbinger and Y.~Mellier,
Astron. Astrophys. \textbf{396}, 1-20 (2002)
doi:10.1051/0004-6361:20021341
[arXiv:astro-ph/0206182 [astro-ph]].

\bibitem{Kaiser:1992ps}
N.~Kaiser and G.~Squires,
Astrophys. J. \textbf{404}, 441-450 (1993)
doi:10.1086/172297

\bibitem{Fahlman:1994np}
G.~Fahlman, N.~Kaiser, G.~Squires and D.~Woods,
Astrophys. J. \textbf{437}, 56-62 (1994)
doi:10.1086/174974
[arXiv:astro-ph/9402017 [astro-ph]].

\bibitem{Seitz:1995dq}
S.~Seitz and P.~Schneider,
Astron. Astrophys. \textbf{305}, 383 (1996)
[arXiv:astro-ph/9503096 [astro-ph]].

\bibitem{Schneider:1996ug}
P.~Schneider,
Mon. Not. Roy. Astron. Soc. \textbf{283}, 837-853 (1996)
doi:10.1093/mnras/283.3.837
[arXiv:astro-ph/9601039 [astro-ph]].





\bibitem{Hamana:2003ts}
T.~Hamana, M.~Takada and N.~Yoshida,
Mon. Not. Roy. Astron. Soc. \textbf{350}, 893 (2004)
doi:10.1111/j.1365-2966.2004.07691.x
[arXiv:astro-ph/0310607 [astro-ph]].

\bibitem{Bernardeau:2000et}
F.~Bernardeau and P.~Valageas,
Astron. Astrophys. \textbf{364}, 1 (2000)
[arXiv:astro-ph/0006270 [astro-ph]].

\bibitem{Liu:2018dsw}
J.~Liu and M.~S.~Madhavacheril,
Phys. Rev. D \textbf{99}, no.8, 083508 (2019)
doi:10.1103/PhysRevD.99.083508
[arXiv:1809.10747 [astro-ph.CO]].

\bibitem{Barthelemy:2020yva}
A.~Barthelemy, S.~Codis and F.~Bernardeau,
Mon. Not. Roy. Astron. Soc. \textbf{503}, no.4, 5204-5222 (2021)
doi:10.1093/mnras/stab818
[arXiv:2012.03831 [astro-ph.CO]].

\bibitem{Thiele:2020rig}
L.~Thiele, J.~C.~Hill and K.~M.~Smith,
Phys. Rev. D \textbf{102}, no.12, 123545 (2020)
doi:10.1103/PhysRevD.102.123545
[arXiv:2009.06547 [astro-ph.CO]].

\bibitem{Boyle:2020bqn}
A.~Boyle, C.~Uhlemann, O.~Friedrich, A.~Barthelemy, S.~Codis, F.~Bernardeau, C.~Giocoli and M.~Baldi,
Mon. Not. Roy. Astron. Soc. \textbf{505}, no.2, 2886-2902 (2021)
doi:10.1093/mnras/stab1381
[arXiv:2012.07771 [astro-ph.CO]].

\bibitem{Giblin:2022ucn}
B.~Giblin, Y.~C.~Cai and J.~Harnois-D{\'e}raps,
Mon. Not. Roy. Astron. Soc. \textbf{520}, no.2, 1721-1737 (2023)
doi:10.1093/mnras/stad230
[arXiv:2211.05708 [astro-ph.CO]].

\bibitem{DES:2017eav}
D.~Gruen \textit{et al.} [DES],
Phys. Rev. D \textbf{98}, no.2, 023507 (2018)
doi:10.1103/PhysRevD.98.023507
[arXiv:1710.05045 [astro-ph.CO]].

\bibitem{DES:2017hhj}
O.~Friedrich \textit{et al.} [DES],
Phys. Rev. D \textbf{98}, no.2, 023508 (2018)
doi:10.1103/PhysRevD.98.023508
[arXiv:1710.05162 [astro-ph.CO]].

\bibitem{Thiele:2023gqr}
L.~Thiele, G.~A.~Marques, J.~Liu and M.~Shirasaki,
Phys. Rev. D \textbf{108}, no.12, 123526 (2023)
doi:10.1103/PhysRevD.108.123526
[arXiv:2304.05928 [astro-ph.CO]].

\bibitem{Takada:2002ee}
M.~Takada and B.~Jain,
Mon. Not. Roy. Astron. Soc. \textbf{337}, 875-894 (2002)
doi:10.1046/j.1365-8711.2002.05972.x
[arXiv:astro-ph/0205055 [astro-ph]].

\bibitem{Jarvis:2003wq}
M.~Jarvis, G.~Bernstein and B.~Jain,
Mon. Not. Roy. Astron. Soc. \textbf{352}, 338-352 (2004)
doi:10.1111/j.1365-2966.2004.07926.x
[arXiv:astro-ph/0307393 [astro-ph]].

\bibitem{DES:2021lsy}
M.~Gatti \textit{et al.} [DES],
Phys. Rev. D \textbf{106}, no.8, 083509 (2022)
doi:10.1103/PhysRevD.106.083509
[arXiv:2110.10141 [astro-ph.CO]].

\bibitem{DES:2025llp}
R.~C.~H.~Gomes \textit{et al.} [DES],
Phys. Rev. D \textbf{112}, no.12, 123515 (2025)
doi:10.1103/sxlz-t9gb
[arXiv:2508.14018 [astro-ph.CO]].

\bibitem{Liu:2014fzc}
J.~Liu, A.~Petri, Z.~Haiman, L.~Hui, J.~M.~Kratochvil and M.~May,
Phys. Rev. D \textbf{91}, no.6, 063507 (2015)
doi:10.1103/PhysRevD.91.063507
[arXiv:1412.0757 [astro-ph.CO]].

\bibitem{Hamana:2015bwa}
T.~Hamana, J.~Sakurai, M.~Koike and L.~Miller,
Publ. Astron. Soc. Jap. \textbf{67}, no.3, 34 (2015)
doi:10.1093/pasj/psv034
[arXiv:1503.01851 [astro-ph.CO]].

\bibitem{DES:2016jfa}
T.~Kacprzak \textit{et al.} [DES],
Mon. Not. Roy. Astron. Soc. \textbf{463}, no.4, 3653-3673 (2016)
doi:10.1093/mnras/stw2070
[arXiv:1603.05040 [astro-ph.CO]].

\bibitem{Shan:2017mgz}
H.~Shan, X.~Liu, H.~Hildebrandt, C.~Pan, N.~Martinet, Z.~Fan, P.~Schneider, M.~Asgari, J.~Harnois-D{\'e}raps and H.~Hoekstra, \textit{et al.}
Mon. Not. Roy. Astron. Soc. \textbf{474}, no.1, 1116-1134 (2018)
doi:10.1093/mnras/stx2837
[arXiv:1709.07651 [astro-ph.CO]].

\bibitem{Martinet:2017rqp}
N.~Martinet, P.~Schneider, H.~Hildebrandt, H.~Shan, M.~Asgari, J.~P.~Dietrich, J.~Harnois-D{\'e}raps, T.~Erben, A.~Grado and C.~Heymans, \textit{et al.}
Mon. Not. Roy. Astron. Soc. \textbf{474}, no.1, 712-730 (2018)
doi:10.1093/mnras/stx2793
[arXiv:1709.07678 [astro-ph.CO]].

\bibitem{Harnois-Deraps:2020pvj}
J.~Harnois-D{\'e}raps, N.~Martinet, T.~Castro, K.~Dolag, B.~Giblin, C.~Heymans, H.~Hildebrandt and Q.~Xia,
Mon. Not. Roy. Astron. Soc. \textbf{506}, no.2, 1623-1650 (2021)
doi:10.1093/mnras/stab1623
[arXiv:2012.02777 [astro-ph.CO]].

\bibitem{DES:2021epj}
D.~Z{\"u}rcher \textit{et al.} [DES],
Mon. Not. Roy. Astron. Soc. \textbf{511}, no.2, 2075-2104 (2022)
doi:10.1093/mnras/stac078
[arXiv:2110.10135 [astro-ph.CO]].

\bibitem{Liu:2022gnc}
X.~Liu, S.~Yuan, C.~Pan, T.~Zhang, Q.~Wang and Z.~Fan,
Mon. Not. Roy. Astron. Soc. \textbf{519}, no.1, 594-612 (2022)
doi:10.1093/mnras/stac2971
[arXiv:2210.07853 [astro-ph.CO]].

\bibitem{Marques:2023bnr}
G.~A.~Marques, J.~Liu, M.~Shirasaki, L.~Thiele, D.~Grand{\'o}n, K.~M.~Huffenberger, S.~Cheng, J.~Harnois-D{\'e}raps, K.~Osato and W.~R.~Coulton,
Mon. Not. Roy. Astron. Soc. \textbf{528}, no.3, 4513-4527 (2024)
doi:10.1093/mnras/stae098
[arXiv:2308.10866 [astro-ph.CO]].

\bibitem{Hennawi:2004ai}
J.~F.~Hennawi and D.~N.~Spergel,
Astrophys. J. \textbf{624}, 59 (2005)
doi:10.1086/428749
[arXiv:astro-ph/0404349 [astro-ph]].

\bibitem{Maturi:2004rn}
M.~Maturi, M.~Meneghetti, M.~Bartelmann, K.~Dolag and L.~Moscardini,
Astron. Astrophys. \textbf{442}, 851 (2005)
doi:10.1051/0004-6361:20042600
[arXiv:astro-ph/0412604 [astro-ph]].

\bibitem{Marian:2011rg}
L.~Marian, R.~E.~Smith, S.~Hilbert and P.~Schneider,
Mon. Not. Roy. Astron. Soc. \textbf{423}, 1711 (2012)
doi:10.1111/j.1365-2966.2012.20992.x
[arXiv:1110.4635 [astro-ph.CO]].

\bibitem{Hamana:2012ur}
T.~Hamana, M.~Oguri, M.~Shirasaki and M.~Sato,
Mon. Not. Roy. Astron. Soc. \textbf{425}, 2287-2298 (2012)
doi:10.1111/j.1365-2966.2012.21582.x
[arXiv:1204.6117 [astro-ph.CO]].

\bibitem{Yang:2011zzn}
X.~Yang, J.~M.~Kratochvil, S.~Wang, E.~A.~Lim, Z.~Haiman and M.~May,
Phys. Rev. D \textbf{84}, 043529 (2011)
doi:10.1103/PhysRevD.84.043529
[arXiv:1109.6333 [astro-ph.CO]].

\bibitem{Liu:2016xjb}
J.~Liu and Z.~Haiman,
Phys. Rev. D \textbf{94}, no.4, 043533 (2016)
doi:10.1103/PhysRevD.94.043533
[arXiv:1606.01318 [astro-ph.CO]].

\bibitem{Sabyr:2021vpr}
A.~Sabyr, Z.~Haiman, J.~M.~Z.~Matilla and T.~Lu,
Phys. Rev. D \textbf{105}, no.2, 023505 (2022)
doi:10.1103/PhysRevD.105.023505
[arXiv:2109.00547 [astro-ph.CO]].

\bibitem{Matsubara:2003yt}
T.~Matsubara,
Astrophys. J. \textbf{584}, 1-33 (2003)
doi:10.1086/345521

\bibitem{Munshi:2011wu}
D.~Munshi, L.~van Waerbeke, J.~Smidt and P.~Coles,
Mon. Not. Roy. Astron. Soc. \textbf{419}, 536 (2012)
doi:10.1111/j.1365-2966.2011.19718.x
[arXiv:1103.1876 [astro-ph.CO]].

\bibitem{Matsubara:2020fet}
T.~Matsubara and S.~Kuriki,
Phys. Rev. D \textbf{104}, no.10, 103522 (2021)
doi:10.1103/PhysRevD.104.103522
[arXiv:2011.04954 [astro-ph.CO]].

\bibitem{Petri:2013ffb}
A.~Petri, Z.~Haiman, L.~Hui, M.~May and J.~M.~Kratochvil,
Phys. Rev. D \textbf{88}, no.12, 123002 (2013)
doi:10.1103/PhysRevD.88.123002
[arXiv:1309.4460 [astro-ph.CO]].

\bibitem{Kratochvil:2011eh}
J.~M.~Kratochvil, E.~A.~Lim, S.~Wang, Z.~Haiman, M.~May and K.~Huffenberger,
Phys. Rev. D \textbf{85}, 103513 (2012)
doi:10.1103/PhysRevD.85.103513
[arXiv:1109.6334 [astro-ph.CO]].

\bibitem{Shirasaki:2013zpa}
M.~Shirasaki and N.~Yoshida,
Astrophys. J. \textbf{786}, 43 (2014)
doi:10.1088/0004-637X/786/1/43
[arXiv:1312.5032 [astro-ph.CO]].

\bibitem{Armijo:2024ujo}
J.~Armijo, G.~A.~Marques, C.~P.~Novaes, L.~Thiele, J.~A.~Cowell, D.~Grand{\'o}n, M.~Shirasaki and J.~Liu,
Mon. Not. Roy. Astron. Soc. \textbf{537}, no.4, 3553-3560 (2025)
doi:10.1093/mnras/staf257
[arXiv:2410.00401 [astro-ph.CO]].

\bibitem{Cheng:2021xdw}
S.~Cheng and B.~M{\'e}nard,
[arXiv:2112.01288 [astro-ph.IM]].

\bibitem{Cheng:2020qbx}
S.~Cheng, Y.~S.~Ting, B.~M{\'e}nard and J.~Bruna,
Mon. Not. Roy. Astron. Soc. \textbf{499}, no.4, 5902-5914 (2020)
doi:10.1093/mnras/staa3165
[arXiv:2006.08561 [astro-ph.CO]].

\bibitem{Ribli:2019wtw}
D.~Ribli, B.~{\'A}.~Pataki, J.~M.~Zorrilla Matilla, D.~Hsu, Z.~Haiman and I.~Csabai,
Mon. Not. Roy. Astron. Soc. \textbf{490}, no.2, 1843-1860 (2019)
doi:10.1093/mnras/stz2610
[arXiv:1902.03663 [astro-ph.CO]].

\bibitem{Cheng:2024kjv}
S.~Cheng, G.~A.~Marques, D.~Grand{\'o}n, L.~Thiele, M.~Shirasaki, B.~M{\'e}nard and J.~Liu,
JCAP \textbf{01}, 006 (2025)
doi:10.1088/1475-7516/2025/01/006
[arXiv:2404.16085 [astro-ph.CO]].

\bibitem{Newman:2022rbn}
J.~A.~Newman and D.~Gruen,
Ann. Rev. Astron. Astrophys. \textbf{60}, 363-414 (2022)
doi:10.1146/annurev-astro-032122-014611
[arXiv:2206.13633 [astro-ph.CO]].

\bibitem{Smith:2002dz}
R.~E.~Smith \textit{et al.} [VIRGO Consortium],
Mon. Not. Roy. Astron. Soc. \textbf{341}, 1311 (2003)
doi:10.1046/j.1365-8711.2003.06503.x
[arXiv:astro-ph/0207664 [astro-ph]].

\bibitem{Heitmann:2008eq}
K.~Heitmann, M.~White, C.~Wagner, S.~Habib and D.~Higdon,
Astrophys. J. \textbf{715}, 104-121 (2010)
doi:10.1088/0004-637X/715/1/104
[arXiv:0812.1052 [astro-ph]].

\bibitem{Heitmann:2009cu}
K.~Heitmann, D.~Higdon, M.~White, S.~Habib, B.~J.~Williams and C.~Wagner,
Astrophys. J. \textbf{705}, 156-174 (2009)
doi:10.1088/0004-637X/705/1/156
[arXiv:0902.0429 [astro-ph.CO]].

\bibitem{Lawrence:2009uk}
E.~Lawrence, K.~Heitmann, M.~White, D.~Higdon, C.~Wagner, S.~Habib and B.~Williams,
Astrophys. J. \textbf{713}, 1322-1331 (2010)
doi:10.1088/0004-637X/713/2/1322
[arXiv:0912.4490 [astro-ph.CO]].

\bibitem{Agarwal:2012ew}
S.~Agarwal, F.~B.~Abdalla, H.~A.~Feldman, O.~Lahav and S.~A.~Thomas,
Mon. Not. Roy. Astron. Soc. \textbf{424}, 1409-1418 (2012)
doi:10.1111/j.1365-2966.2012.21326.x
[arXiv:1203.1695 [astro-ph.CO]].

\bibitem{Takahashi:2012em}
R.~Takahashi, M.~Sato, T.~Nishimichi, A.~Taruya and M.~Oguri,
Astrophys. J. \textbf{761}, 152 (2012)
doi:10.1088/0004-637X/761/2/152
[arXiv:1208.2701 [astro-ph.CO]].

\bibitem{Agarwal:2013aea}
S.~Agarwal, F.~B.~Abdalla, H.~A.~Feldman, O.~Lahav and S.~A.~Thomas,
Mon. Not. Roy. Astron. Soc. \textbf{439}, no.2, 2102-2121 (2014)
doi:10.1093/mnras/stu090
[arXiv:1312.2101 [astro-ph.CO]].

\bibitem{Euclid:2018mlb}
M.~Knabenhans \textit{et al.} [Euclid],
Mon. Not. Roy. Astron. Soc. \textbf{484}, 5509-5529 (2019)
doi:10.1093/mnras/stz197
[arXiv:1809.04695 [astro-ph.CO]].

\bibitem{Angulo:2020vky}
R.~E.~Angulo, M.~Zennaro, S.~Contreras, G.~Aric{\`o}, M.~Pellejero-Iba{\~n}ez and J.~St{\"u}cker,
Mon. Not. Roy. Astron. Soc. \textbf{507}, no.4, 5869-5881 (2021)
doi:10.1093/mnras/stab2018
[arXiv:2004.06245 [astro-ph.CO]].

\bibitem{Moran:2022iwe}
K.~R.~Moran, K.~Heitmann, E.~Lawrence, S.~Habib, D.~Bingham, A.~Upadhye, J.~Kwan, D.~Higdon and R.~Payne,
Mon. Not. Roy. Astron. Soc. \textbf{520}, no.3, 3443-3458 (2023)
doi:10.1093/mnras/stac3452
[arXiv:2207.12345 [astro-ph.CO]].

\bibitem{Chen:2025ugn}
Z.~Chen, Y.~Yu, J.~Han and Y.~Jing,
Sci. China Phys. Mech. Astron. \textbf{68}, no.8, 289512 (2025)
doi:10.1007/s11433-025-2671-0
[arXiv:2502.11160 [astro-ph.CO]].

\bibitem{Rudd:2007zx}
D.~H.~Rudd, A.~R.~Zentner and A.~V.~Kravtsov,
Astrophys. J. \textbf{672}, 19-32 (2008)
doi:10.1086/523836
[arXiv:astro-ph/0703741 [astro-ph]].

\bibitem{vanDaalen:2011xb}
M.~P.~van Daalen, J.~Schaye, C.~M.~Booth and C.~D.~Vecchia,
Mon. Not. Roy. Astron. Soc. \textbf{415}, 3649-3665 (2011)
doi:10.1111/j.1365-2966.2011.18981.x
[arXiv:1104.1174 [astro-ph.CO]].

\bibitem{Fedeli:2014gja}
C.~Fedeli, E.~Semboloni, M.~Velliscig, M.~Van Daalen, J.~Schaye and H.~Hoekstra,
JCAP \textbf{08}, 028 (2014)
doi:10.1088/1475-7516/2014/08/028
[arXiv:1406.5013 [astro-ph.CO]].

\bibitem{Osato:2015lja}
K.~Osato, M.~Shirasaki and N.~Yoshida,
Astrophys. J. \textbf{806}, no.2, 186 (2015)
doi:10.1088/0004-637X/806/2/186
[arXiv:1501.02055 [astro-ph.CO]].

\bibitem{Mead:2015yca}
A.~Mead, J.~Peacock, C.~Heymans, S.~Joudaki and A.~Heavens,
Mon. Not. Roy. Astron. Soc. \textbf{454}, no.2, 1958-1975 (2015)
doi:10.1093/mnras/stv2036
[arXiv:1505.07833 [astro-ph.CO]].

\bibitem{Mead:2016zqy}
A.~Mead, C.~Heymans, L.~Lombriser, J.~Peacock, O.~Steele and H.~Winther,
Mon. Not. Roy. Astron. Soc. \textbf{459}, no.2, 1468-1488 (2016)
doi:10.1093/mnras/stw681
[arXiv:1602.02154 [astro-ph.CO]].

\bibitem{Chisari:2018prw}
N.~E.~Chisari, M.~L.~A.~Richardson, J.~Devriendt, Y.~Dubois, A.~Schneider, A.~L.~Brun, M.C., R.~S.~Beckmann, S.~Peirani, A.~Slyz and C.~Pichon,
Mon. Not. Roy. Astron. Soc. \textbf{480}, no.3, 3962-3977 (2018)
doi:10.1093/mnras/sty2093
[arXiv:1801.08559 [astro-ph.CO]].

\bibitem{Barreira:2019ckp}
A.~Barreira, D.~Nelson, A.~Pillepich, V.~Springel, F.~Schmidt, R.~Pakmor, L.~Hernquist and M.~Vogelsberger,
Mon. Not. Roy. Astron. Soc. \textbf{488}, no.2, 2079-2092 (2019)
doi:10.1093/mnras/stz1807
[arXiv:1904.02070 [astro-ph.CO]].

\bibitem{Chisari:2019tus}
N.~E.~Chisari, A.~J.~Mead, S.~Joudaki, P.~Ferreira, A.~Schneider, J.~Mohr, T.~Tr{\"o}ster, D.~Alonso, I.~G.~McCarthy and S.~Martin-Alvarez, \textit{et al.}
Open J. Astrophys. \textbf{2}, no.1, 4 (2019)
doi:10.21105/astro.1905.06082
[arXiv:1905.06082 [astro-ph.CO]].

\bibitem{Arico:2019ykw}
G.~Aric{\`o}, R.~E.~Angulo, C.~Hern{\'a}ndez-Monteagudo, S.~Contreras, M.~Zennaro, M.~Pellejero-Iba{\~n}ez and Y.~Rosas-Guevara,
Mon. Not. Roy. Astron. Soc. \textbf{495}, no.4, 4800-4819 (2020)
doi:10.1093/mnras/staa1478
[arXiv:1911.08471 [astro-ph.CO]].

\bibitem{Mead:2020vgs}
A.~Mead, S.~Brieden, T.~Tr{\"o}ster and C.~Heymans,
Mon. Not. Roy. Astron. Soc. \textbf{502}, no.1, 1401-1422 (2021)
doi:10.1093/mnras/stab082
[arXiv:2009.01858 [astro-ph.CO]].

\bibitem{Osato:2020sxo}
K.~Osato, J.~Liu and Z.~Haiman,
Mon. Not. Roy. Astron. Soc. \textbf{502}, no.4, 5593-5602 (2021)
doi:10.1093/mnras/stab395
[arXiv:2010.09731 [astro-ph.CO]].

\bibitem{Giri:2021qin}
S.~K.~Giri and A.~Schneider,
JCAP \textbf{12}, no.12, 046 (2021)
doi:10.1088/1475-7516/2021/12/046
[arXiv:2108.08863 [astro-ph.CO]].

\bibitem{Acuto:2021yjm}
A.~Acuto, I.~G.~McCarthy, J.~Kwan, J.~Salcido, S.~G.~Stafford and A.~S.~Font,
Mon. Not. Roy. Astron. Soc. \textbf{508}, no.3, 3519-3534 (2021)
doi:10.1093/mnras/stab2834
[arXiv:2109.11855 [astro-ph.CO]].

\bibitem{Salcido:2023etz}
J.~Salcido, I.~G.~McCarthy, J.~Kwan, A.~Upadhye and A.~S.~Font,
Mon. Not. Roy. Astron. Soc. \textbf{523}, no.2, 2247-2262 (2023)
doi:10.1093/mnras/stad1474
[arXiv:2305.09710 [astro-ph.CO]].

\bibitem{Schaller:2024jiq}
M.~Schaller, J.~Schaye, R.~Kugel, J.~C.~Broxterman and M.~P.~van Daalen,
Mon. Not. Roy. Astron. Soc. \textbf{539}, no.2, 1337-1351 (2025)
doi:10.1093/mnras/staf569
[arXiv:2410.17109 [astro-ph.CO]].

\bibitem{Schneider:2025zca}
A.~Schneider, M.~Kova{\v{c}}, J.~Bucko, A.~Nicola, R.~Reischke, S.~K.~Giri, R.~Teyssier, T.~Tr{\"o}ster, A.~Refregier and M.~Schaller, \textit{et al.}
JCAP \textbf{12}, 043 (2025)
doi:10.1088/1475-7516/2025/12/043
[arXiv:2507.07892 [astro-ph.CO]].

\bibitem{Kovac:2025zqy}
M.~Kova{\v{c}}, A.~Nicola, J.~Bucko, A.~Schneider, R.~Reischke, S.~K.~Giri, R.~Teyssier, M.~Schaller and J.~Schaye,
JCAP \textbf{11}, 046 (2025)
doi:10.1088/1475-7516/2025/11/046
[arXiv:2507.07991 [astro-ph.CO]].

\bibitem{Hamana:2001kd}
T.~Hamana,
Mon. Not. Roy. Astron. Soc. \textbf{326}, 326 (2001)
doi:10.1046/j.1365-8711.2001.04607.x
[arXiv:astro-ph/0104244 [astro-ph]].

\bibitem{Schmidt:2009rh}
F.~Schmidt, E.~Rozo, S.~Dodelson, L.~Hui and E.~Sheldon,
Phys. Rev. Lett. \textbf{103}, 051301 (2009)
doi:10.1103/PhysRevLett.103.051301
[arXiv:0904.4702 [astro-ph.CO]].

\bibitem{Schmidt:2009ri}
F.~Schmidt, E.~Rozo, S.~Dodelson, L.~Hui and E.~Sheldon,
Astrophys. J. \textbf{702}, 593-602 (2009)
doi:10.1088/0004-637X/702/1/593
[arXiv:0904.4703 [astro-ph.CO]].

\bibitem{Krause:2009yr}
E.~Krause and C.~M.~Hirata,
Astron. Astrophys. \textbf{523}, A28 (2010)
doi:10.1051/0004-6361/200913524
[arXiv:0910.3786 [astro-ph.CO]].

\bibitem{Schmidt:2010ex}
F.~Schmidt and E.~Rozo,
Astrophys. J. \textbf{735}, 119 (2011)
doi:10.1088/0004-637X/735/2/119
[arXiv:1009.0757 [astro-ph.CO]].

\bibitem{Liu:2013yna}
J.~Liu, Z.~Haiman, L.~Hui, J.~M.~Kratochvil and M.~May,
Phys. Rev. D \textbf{89}, no.2, 023515 (2014)
doi:10.1103/PhysRevD.89.023515
[arXiv:1310.7517 [astro-ph.CO]].

\bibitem{Deshpande:2019sdl}
A.~C.~Deshpande, T.~D.~Kitching, V.~F.~Cardone, P.~L.~Taylor, S.~Casas, S.~Camera, C.~Carbone, M.~Kilbinger, V.~Pettorino and Z.~Sakr, \textit{et al.}
Astron. Astrophys. \textbf{636}, A95 (2020)
doi:10.1051/0004-6361/201937323
[arXiv:1912.07326 [astro-ph.CO]].

\bibitem{Duncan:2021jxl}
C.~A.~J.~Duncan, J.~Harnois-D{\'e}raps, L.~Miller and A.~Langedijk,
Mon. Not. Roy. Astron. Soc. \textbf{515}, no.1, 1130-1145 (2022)
doi:10.1093/mnras/stac1809
[arXiv:2111.09867 [astro-ph.CO]].

\bibitem{Hamana:2000wb}
T.~Hamana, S.~T.~Colombi, A.~Thion, J.~E.~G.~T.~Devriendt, Y.~Mellier and F.~Bernardeau,
Mon. Not. Roy. Astron. Soc. \textbf{330}, 365 (2002)
doi:10.1046/j.1365-8711.2002.05103.x
[arXiv:astro-ph/0012200 [astro-ph]].

\bibitem{Schneider:2001af}
P.~Schneider, L.~Van Waerbeke and Y.~Mellier,
Astron. Astrophys. \textbf{389}, 729-741 (2002)
doi:10.1051/0004-6361:20020626
[arXiv:astro-ph/0112441 [astro-ph]].

\bibitem{Valageas:2013qfa}
P.~Valageas,
Astron. Astrophys. \textbf{561}, A53 (2014)
doi:10.1051/0004-6361/201322146
[arXiv:1306.6151 [astro-ph.CO]].

\bibitem{Yu:2014iea}
Y.~Yu, P.~Zhang, W.~Lin and W.~Cui,
Astrophys. J. \textbf{803}, no.1, 46 (2015)
doi:10.1088/0004-637X/803/1/46
[arXiv:1402.7144 [astro-ph.CO]].

\bibitem{DES:2023ycm}
M.~Gatti \textit{et al.} [DES],
Mon. Not. Roy. Astron. Soc. \textbf{527}, no.1, L115-L121 (2024)
doi:10.1093/mnrasl/slad143
[arXiv:2307.13860 [astro-ph.CO]].

\bibitem{KiDS-1000:2024rdr}
L.~Linke \textit{et al.} [KiDS-1000 and Euclid],
Astron. Astrophys. \textbf{693}, A210 (2025)
doi:10.1051/0004-6361/202451494
[arXiv:2407.09810 [astro-ph.CO]].

\bibitem{Duncan:2024dzj}
C.~A.~J.~Duncan and M.~L.~Brown,
Mon. Not. Roy. Astron. Soc. \textbf{541}, no.4, 3549-3560 (2025)
doi:10.1093/mnras/staf1148
[arXiv:2411.15063 [astro-ph.CO]].

\bibitem{Troxel:2014dba}
M.~A.~Troxel and M.~Ishak,
Phys. Rept. \textbf{558}, 1-59 (2014)
doi:10.1016/j.physrep.2014.11.001
[arXiv:1407.6990 [astro-ph.CO]].

\bibitem{Chisari:2025gsy}
N.~E.~Chisari,
Astron. Astrophys. Rev. \textbf{33}, no.1, 5 (2025)
doi:10.1007/s00159-025-00161-8
[arXiv:2510.15738 [astro-ph.CO]].

\bibitem{Friedrich:2015nga}
O.~Friedrich, S.~Seitz, T.~F.~Eifler and D.~Gruen,
Mon. Not. Roy. Astron. Soc. \textbf{456}, no.3, 2662-2680 (2016)
doi:10.1093/mnras/stv2833
[arXiv:1508.00895 [astro-ph.CO]].

\bibitem{Shirasaki:2016fuf}
M.~Shirasaki, M.~Takada, H.~Miyatake, R.~Takahashi, T.~Hamana, T.~Nishimichi and R.~Murata,
Mon. Not. Roy. Astron. Soc. \textbf{470}, no.3, 3476-3496 (2017)
doi:10.1093/mnras/stx1477
[arXiv:1607.08679 [astro-ph.CO]].

\bibitem{Hartlap:2006kj}
J.~Hartlap, P.~Simon and P.~Schneider,
Astron. Astrophys. \textbf{464}, 399 (2007)
doi:10.1051/0004-6361:20066170
[arXiv:astro-ph/0608064 [astro-ph]].

\bibitem{Taylor:2012kz}
A.~Taylor, B.~Joachimi and T.~Kitching,
Mon. Not. Roy. Astron. Soc. \textbf{432}, 1928 (2013)
doi:10.1093/mnras/stt270
[arXiv:1212.4359 [astro-ph.CO]].

\bibitem{Dodelson:2013uaa}
S.~Dodelson and M.~D.~Schneider,
Phys. Rev. D \textbf{88}, 063537 (2013)
doi:10.1103/PhysRevD.88.063537
[arXiv:1304.2593 [astro-ph.CO]].

\bibitem{Takahashi:2017hjr}
R.~Takahashi, T.~Hamana, M.~Shirasaki, T.~Namikawa, T.~Nishimichi, K.~Osato and K.~Shiroyama,
Astrophys. J. \textbf{850}, no.1, 24 (2017)
doi:10.3847/1538-4357/aa943d
[arXiv:1706.01472 [astro-ph.CO]].

\bibitem{Shirasaki:2019gya}
M.~Shirasaki, T.~Hamana, M.~Takada, R.~Takahashi and H.~Miyatake,
Mon. Not. Roy. Astron. Soc. \textbf{486}, no.1, 52-69 (2019)
doi:10.1093/mnras/stz791
[arXiv:1901.09488 [astro-ph.CO]].

\bibitem{Sato:2009ct}
M.~Sato, T.~Hamana, R.~Takahashi, M.~Takada, N.~Yoshida, T.~Matsubara and N.~Sugiyama,
Astrophys. J. \textbf{701}, 945-954 (2009)
doi:10.1088/0004-637X/701/2/945
[arXiv:0906.2237 [astro-ph.CO]].

\bibitem{Pielorz:2009sv}
J.~Pielorz, J.~Rodiger, I.~Tereno and P.~Schneider,
Astron. Astrophys. \textbf{514}, A79 (2010)
doi:10.1051/0004-6361/200912854
[arXiv:0907.1524 [astro-ph.CO]].

\bibitem{Takada:2013wfa}
M.~Takada and W.~Hu,
Phys. Rev. D \textbf{87}, no.12, 123504 (2013)
doi:10.1103/PhysRevD.87.123504
[arXiv:1302.6994 [astro-ph.CO]].

\bibitem{Krause:2016jvl}
E.~Krause and T.~Eifler,
Mon. Not. Roy. Astron. Soc. \textbf{470}, no.2, 2100-2112 (2017)
doi:10.1093/mnras/stx1261
[arXiv:1601.05779 [astro-ph.CO]].

\bibitem{Mohammed:2016sre}
I.~Mohammed, U.~Seljak and Z.~Vlah,
Mon. Not. Roy. Astron. Soc. \textbf{466}, no.1, 780-797 (2017)
doi:10.1093/mnras/stw3196
[arXiv:1607.00043 [astro-ph.CO]].


\bibitem{Barreira:2017fjz}
A.~Barreira, E.~Krause and F.~Schmidt,
JCAP \textbf{06}, 015 (2018)
doi:10.1088/1475-7516/2018/06/015
[arXiv:1711.07467 [astro-ph.CO]].

\bibitem{Kayo:2012nm}
I.~Kayo, M.~Takada and B.~Jain,
Mon. Not. Roy. Astron. Soc. \textbf{429}, 344-371 (2013)
doi:10.1093/mnras/sts340
[arXiv:1207.6322 [astro-ph.CO]].

\bibitem{Shirasaki:2015dga}
M.~Shirasaki, T.~Hamana and N.~Yoshida,
Mon. Not. Roy. Astron. Soc. \textbf{453}, no.3, 3043-3067 (2015)
doi:10.1093/mnras/stv1854
[arXiv:1504.05672 [astro-ph.CO]].

\bibitem{Uhlemann:2022znd}
C.~Uhlemann, O.~Friedrich, A.~Boyle, A.~Gough, A.~Barthelemy, F.~Bernardeau and S.~Codis,
Open J. Astrophys. \textbf{6}, 2023
doi:10.21105/astro.2210.07819
[arXiv:2210.07819 [astro-ph.CO]].

\bibitem{Bayer:2022nws}
A.~E.~Bayer, J.~Liu, R.~Terasawa, A.~Barreira, Y.~Zhong and Y.~Feng,
Phys. Rev. D \textbf{108}, no.4, 043521 (2023)
doi:10.1103/PhysRevD.108.043521
[arXiv:2210.15647 [astro-ph.CO]].



\bibitem{Linke:2022xnl}
L.~Linke, S.~Heydenreich, P.~A.~Burger and P.~Schneider,
Astron. Astrophys. \textbf{672}, A185 (2023)
doi:10.1051/0004-6361/202245652
[arXiv:2212.04485 [astro-ph.CO]].

\bibitem{Wielders:2025ust}
N.~Wielders, L.~Linke, P.~A.~Burger, S.~Heydenreich, L.~Porth and P.~Schneider,
Astron. Astrophys. \textbf{702}, A207 (2025)
doi:10.1051/0004-6361/202554474
[arXiv:2509.20443 [astro-ph.CO]].

\bibitem{Heavens:2017efz}
A.~Heavens, E.~Sellentin, D.~de Mijolla and A.~Vianello,
Mon. Not. Roy. Astron. Soc. \textbf{472}, no.4, 4244-4250 (2017)
doi:10.1093/mnras/stx2326
[arXiv:1707.06529 [astro-ph.CO]].

\bibitem{Bellini:2019lnn}
E.~Bellini, D.~Alonso, S.~Joudaki and L.~van Waerbeke,
doi:10.21105/astro.1903.04957
[arXiv:1903.04957 [astro-ph.CO]].

\bibitem{Heavens:2020spq}
A.~Heavens, E.~Sellentin and A.~Jaffe,
Mon. Not. Roy. Astron. Soc. \textbf{498}, no.3, 3440-3451 (2020)
doi:10.1093/mnras/staa2589
[arXiv:2006.06706 [astro-ph.CO]].

\bibitem{Philcox:2020zyp}
O.~H.~E.~Philcox, M.~M.~Ivanov, M.~Zaldarriaga, M.~Simonovic and M.~Schmittfull,
Phys. Rev. D \textbf{103}, no.4, 043508 (2021)
doi:10.1103/PhysRevD.103.043508
[arXiv:2009.03311 [astro-ph.CO]].

\bibitem{Ferreira:2020bgt}
T.~Ferreira \textit{et al.} [LSST Dark Energy Science],
Phys. Rev. D \textbf{103}, no.10, 103535 (2021)
doi:10.1103/PhysRevD.103.103535
[arXiv:2010.15986 [astro-ph.CO]].



\bibitem{Heavens:2023liw}
A.~F.~Heavens, A.~Mootoovaloo, R.~Trotta and E.~Sellentin,
JCAP \textbf{11}, 048 (2023)
doi:10.1088/1475-7516/2023/11/048
[arXiv:2306.15998 [astro-ph.IM]].

\bibitem{Sugiyama:2025bhl}
S.~Sugiyama and M.~Park,
Phys. Rev. D \textbf{112}, no.12, 123505 (2025)
doi:10.1103/hksr-253f
[arXiv:2508.14021 [astro-ph.CO]].

\bibitem{Schneider:2011wf}
M.~D.~Schneider, S.~Cole, C.~S.~Frenk and I.~Szapudi,
Astrophys. J. \textbf{737}, 11 (2011)
doi:10.1088/0004-637X/737/1/11
[arXiv:1103.2767 [astro-ph.CO]].

\bibitem{Morrison:2013tqa}
C.~B.~Morrison and M.~D.~Schneider,
JCAP \textbf{11}, 009 (2013)
doi:10.1088/1475-7516/2013/11/009
[arXiv:1304.7789 [astro-ph.CO]].

\bibitem{Petri:2016wlu}
A.~Petri, Z.~Haiman and M.~May,
Phys. Rev. D \textbf{93}, no.6, 063524 (2016)
doi:10.1103/PhysRevD.93.063524
[arXiv:1601.06792 [astro-ph.CO]].

\bibitem{Friedrich:2017xxx}
O.~Friedrich and T.~Eifler,
Mon. Not. Roy. Astron. Soc. \textbf{473}, no.3, p.4150-4163 (2018)
doi:10.1093/mnras/stx2566
[arXiv:arXiv:1703.07786 [astro-ph.IM]].

\bibitem{Hall:2018umb}
A.~Hall and A.~Taylor,
Mon. Not. Roy. Astron. Soc. \textbf{483}, no.1, 189-207 (2019)
doi:10.1093/mnras/sty3102
[arXiv:1807.06875 [astro-ph.CO]].

\bibitem{Looijmans:2024sna}
M.~J.~Looijmans, M.~Wang and F.~Beutler,
Mon. Not. Roy. Astron. Soc. \textbf{537}, no.1, 21-34 (2025)
doi:10.1093/mnras/stae2786
[arXiv:2402.13783 [astro-ph.CO]].

\bibitem{LSSTDarkEnergyScience:2019hkz}
D.~Korytov \textit{et al.} [LSST Dark Energy Science],
Astrophys. J. Suppl. \textbf{245}, no.2, 26 (2019)
doi:10.3847/1538-4365/ab510c
[arXiv:1907.06530 [astro-ph.CO]].

\bibitem{Abdalla:2022yfr}
E.~Abdalla, G.~Franco Abell{\'a}n, A.~Aboubrahim, A.~Agnello, O.~Akarsu, Y.~Akrami, G.~Alestas, D.~Aloni, L.~Amendola and L.~A.~Anchordoqui, \textit{et al.}
JHEAp \textbf{34}, 49-211 (2022)
doi:10.1016/j.jheap.2022.04.002
[arXiv:2203.06142 [astro-ph.CO]].

\bibitem{KiDS:2020suj}
M.~Asgari \textit{et al.} [KiDS],
Astron. Astrophys. \textbf{645}, A104 (2021)
doi:10.1051/0004-6361/202039070
[arXiv:2007.15633 [astro-ph.CO]].

\bibitem{DES:2021vln}
L.~F.~Secco \textit{et al.} [DES],
Phys. Rev. D \textbf{105}, no.2, 023515 (2022)
doi:10.1103/PhysRevD.105.023515
[arXiv:2105.13544 [astro-ph.CO]].

\bibitem{Dalal:2023olq}
R.~Dalal, X.~Li, A.~Nicola, J.~Zuntz, M.~A.~Strauss, S.~Sugiyama, T.~Zhang, M.~M.~Rau, R.~Mandelbaum and M.~Takada, \textit{et al.}
Phys. Rev. D \textbf{108}, no.12, 123519 (2023)
doi:10.1103/PhysRevD.108.123519
[arXiv:2304.00701 [astro-ph.CO]].

\bibitem{Li:2023tui}
X.~Li, T.~Zhang, S.~Sugiyama, R.~Dalal, R.~Terasawa, M.~M.~Rau, R.~Mandelbaum, M.~Takada, S.~More and M.~A.~Strauss, \textit{et al.}
Phys. Rev. D \textbf{108}, no.12, 123518 (2023)
doi:10.1103/PhysRevD.108.123518
[arXiv:2304.00702 [astro-ph.CO]].

\bibitem{Nishimichi:2020tvu}
T.~Nishimichi, G.~D'Amico, M.~M.~Ivanov, L.~Senatore, M.~Simonovi{\'c}, M.~Takada, M.~Zaldarriaga and P.~Zhang,
Phys. Rev. D \textbf{102}, no.12, 123541 (2020)
doi:10.1103/PhysRevD.102.123541
[arXiv:2003.08277 [astro-ph.CO]].

\bibitem{Dietrich:2009jq}
J.~P.~Dietrich and J.~Hartlap,
Mon. Not. Roy. Astron. Soc. \textbf{402}, 1049 (2010)
doi:10.1111/j.1365-2966.2009.15948.x
[arXiv:0906.3512 [astro-ph.CO]].

\bibitem{Harnois-Deraps:2012ixt}
J.~Harnois-Deraps, S.~Vafaei and L.~Van Waerbeke,
Mon. Not. Roy. Astron. Soc. \textbf{426}, 1262 (2012)
doi:10.1111/j.1365-2966.2012.21624.x
[arXiv:1202.2332 [astro-ph.CO]].

\bibitem{Liu:2017now}
J.~Liu, S.~Bird, J.~M.~Z.~Matilla, J.~C.~Hill, Z.~Haiman, M.~S.~Madhavacheril, A.~Petri and D.~N.~Spergel,
JCAP \textbf{03}, 049 (2018)
doi:10.1088/1475-7516/2018/03/049
[arXiv:1711.10524 [astro-ph.CO]].

\bibitem{Harnois-Deraps:2018bcv}
J.~Harnois-D{\'e}raps, A.~Amon, A.~Choi, V.~Demchenko, C.~Heymans, A.~Kannawadi, R.~Nakajima, E.~Sirks, L.~van Waerbeke and Y.~C.~Cai, \textit{et al.}
Mon. Not. Roy. Astron. Soc. \textbf{481}, no.1, 1337-1367 (2018)
doi:10.1093/mnras/sty2319
[arXiv:1805.04511 [astro-ph.CO]].

\bibitem{Harnois-Deraps:2019rsd}
J.~Harnois-Deraps, B.~Giblin and B.~Joachimi,
Astron. Astrophys. \textbf{631}, A160 (2019)
doi:10.1051/0004-6361/201935912
[arXiv:1905.06454 [astro-ph.CO]].

\bibitem{Shirasaki:2019wxk}
M.~Shirasaki, K.~Moriwaki, T.~Oogi, N.~Yoshida, S.~Ikeda and T.~Nishimichi,
Mon. Not. Roy. Astron. Soc. \textbf{504}, no.2, 1825-1839 (2021)
doi:10.1093/mnras/stab982
[arXiv:1911.12890 [astro-ph.CO]].

\bibitem{Kacprzak:2022pww}
T.~Kacprzak, J.~Fluri, A.~Schneider, A.~Refregier and J.~Stadel,
JCAP \textbf{02}, 050 (2023)
doi:10.1088/1475-7516/2023/02/050
[arXiv:2209.04662 [astro-ph.CO]].

\bibitem{Ferlito:2023gum}
F.~Ferlito, V.~Springel, C.~T.~Davies, C.~Hern{\'a}ndez-Aguayo, R.~Pakmor, M.~Barrera, S.~D.~M.~White, A.~M.~Delgado, B.~Hadzhiyska and L.~Hernquist, \textit{et al.}
Mon. Not. Roy. Astron. Soc. \textbf{524}, no.4, 5591-5606 (2023)
doi:10.1093/mnras/stad2205
[arXiv:2304.12338 [astro-ph.CO]].

\bibitem{DES:2024xij}
N.~Jeffrey \textit{et al.} [DES],
Mon. Not. Roy. Astron. Soc. \textbf{536}, no.2, 1303-1322 (2024)
doi:10.1093/mnras/stae2629
[arXiv:2403.02314 [astro-ph.CO]].

\bibitem{Taruya:2002vy}
A.~Taruya, M.~Takada, T.~Hamana, I.~Kayo and T.~Futamase,
Astrophys. J. \textbf{571}, 638-653 (2002)
doi:10.1086/340048
[arXiv:astro-ph/0202090 [astro-ph]].

\bibitem{Xavier:2016elr}
H.~S.~Xavier, F.~B.~Abdalla and B.~Joachimi,
Mon. Not. Roy. Astron. Soc. \textbf{459}, no.4, 3693-3710 (2016)
doi:10.1093/mnras/stw874
[arXiv:1602.08503 [astro-ph.CO]].

\bibitem{Makiya:2020iai}
R.~Makiya, I.~Kayo and E.~Komatsu,
JCAP \textbf{03}, 095 (2021)
doi:10.1088/1475-7516/2021/03/095
[arXiv:2008.13195 [astro-ph.CO]].

\bibitem{Yu:2016qoq}
Y.~Yu, P.~Zhang and Y.~Jing,
Phys. Rev. D \textbf{94}, no.8, 083520 (2016)
doi:10.1103/PhysRevD.94.083520
[arXiv:1607.05007 [astro-ph.CO]].

\bibitem{Shirasaki:2016vve}
M.~Shirasaki,
Mon. Not. Roy. Astron. Soc. \textbf{465}, no.2, 1974-1983 (2017)
doi:10.1093/mnras/stw2950
[arXiv:1610.00840 [astro-ph.CO]].

\bibitem{Tessore:2023zyk}
N.~Tessore, A.~Loureiro, B.~Joachimi, M.~von Wietersheim-Kramsta and N.~Jeffrey,
doi:10.21105/astro.2302.01942
[arXiv:2302.01942 [astro-ph.CO]].

\bibitem{Zhong:2024wdk}
K.~Zhong, G.~Bernstein, S.~S.~Boruah, B.~Jain and S.~Kobla,
Phys. Rev. D \textbf{111}, no.10, 103507 (2025)
doi:10.1103/PhysRevD.111.103507
[arXiv:2411.04759 [astro-ph.CO]].

\bibitem{Izard:2017kma}
A.~Izard, P.~Fosalba and M.~Crocce,
Mon. Not. Roy. Astron. Soc. \textbf{473}, no.3, 3051-3061 (2018)
doi:10.1093/mnras/stx2544
[arXiv:1707.06312 [astro-ph.CO]].

\bibitem{Sgier:2018soj}
R.~Sgier, A.~R{\'e}fr{\'e}gier, A.~Amara and A.~Nicola,
JCAP \textbf{01}, 044 (2019)
doi:10.1088/1475-7516/2019/01/044
[arXiv:1801.05745 [astro-ph.CO]].

\bibitem{Bohm:2020ilt}
V.~B{\"o}hm, Y.~Feng, M.~E.~Lee and B.~Dai,
Astron. Comput. \textbf{36}, 100490 (2021)
doi:10.1016/j.ascom.2021.100490
[arXiv:2012.07266 [astro-ph.CO]].

\bibitem{Giocoli:2017nnq}
C.~Giocoli, S.~Di Meo, M.~Meneghetti, E.~Jullo, S.~de la Torre, L.~Moscardini, M.~Baldi, P.~Mazzotta and R.~Benton Metcalf,
Mon. Not. Roy. Astron. Soc. \textbf{470}, no.3, 3574-3590 (2017)
doi:10.1093/mnras/stx1399
[arXiv:1701.02739 [astro-ph.CO]].

\bibitem{Giocoli:2020pfb}
C.~Giocoli, P.~Monaco, L.~Moscardini, T.~Castro, M.~Meneghetti, R.~B.~Metcalf and M.~Baldi,
Mon. Not. Roy. Astron. Soc. \textbf{496}, no.2, 1307-1324 (2020)
doi:10.1093/mnras/staa1538
[arXiv:2001.11512 [astro-ph.CO]].

\bibitem{Gupta:2018eev}
A.~Gupta, J.~M.~Z.~Matilla, D.~Hsu and Z.~Haiman,
Phys. Rev. D \textbf{97}, no.10, 103515 (2018)
doi:10.1103/PhysRevD.97.103515
[arXiv:1802.01212 [astro-ph.CO]].

\bibitem{Fluri:2018hoy}
J.~Fluri, T.~Kacprzak, A.~Refregier, A.~Amara, A.~Lucchi and T.~Hofmann,
Phys. Rev. D \textbf{98}, no.12, 123518 (2018)
doi:10.1103/PhysRevD.98.123518
[arXiv:1807.08732 [astro-ph.CO]].

\bibitem{Fluri:2019qtp}
J.~Fluri, T.~Kacprzak, A.~Lucchi, A.~Refregier, A.~Amara, T.~Hofmann and A.~Schneider,
Phys. Rev. D \textbf{100}, no.6, 063514 (2019)
doi:10.1103/PhysRevD.100.063514
[arXiv:1906.03156 [astro-ph.CO]].

\bibitem{Lu:2023dep}
T.~Lu, Z.~Haiman and X.~Li,
Mon. Not. Roy. Astron. Soc. \textbf{521}, no.2, 2050-2066 (2023)
doi:10.1093/mnras/stad686
[arXiv:2301.01354 [astro-ph.CO]].

\bibitem{Ribli:2018kwb}
D.~Ribli, B.~{\'A}.~Pataki and I.~Csabai,
Nature Astron. \textbf{3}, no.1, 93-98 (2019)
doi:10.1038/s41550-018-0596-8
[arXiv:1806.05995 [astro-ph.CO]].

\bibitem{ZorrillaMatilla:2020doz}
J.~M.~Zorrilla Matilla, M.~Sharma, D.~Hsu and Z.~Haiman,
Phys. Rev. D \textbf{102}, no.12, 123506 (2020)
doi:10.1103/PhysRevD.102.123506
[arXiv:2007.06529 [astro-ph.CO]].

\bibitem{Zhong:2024qpf}
K.~Zhong, M.~Gatti and B.~Jain,
Phys. Rev. D \textbf{110}, no.4, 043535 (2024)
doi:10.1103/PhysRevD.110.043535
[arXiv:2403.01368 [astro-ph.CO]].

\bibitem{Sharma:2024pth}
D.~Sharma, B.~Dai and U.~Seljak,
JCAP \textbf{08}, 010 (2024)
doi:10.1088/1475-7516/2024/08/010
[arXiv:2403.03490 [astro-ph.CO]].

\bibitem{Makinen:2024bff}
T.~L.~Makinen, A.~Heavens, N.~Porqueres, T.~Charnock, A.~Lapel and B.~D.~Wandelt,
JCAP \textbf{01}, 095 (2025)
doi:10.1088/1475-7516/2025/01/095
[arXiv:2407.18909 [astro-ph.CO]].

\bibitem{Jeffrey:2020xve}
N.~Jeffrey, J.~Alsing and F.~Lanusse,
Mon. Not. Roy. Astron. Soc. \textbf{501}, no.1, 954-969 (2021)
doi:10.1093/mnras/staa3594
[arXiv:2009.08459 [astro-ph.CO]].

\bibitem{Lin:2022ayr}
K.~Lin, M.~von Wietersheim-Kramsta, B.~Joachimi and S.~Feeney,
Mon. Not. Roy. Astron. Soc. \textbf{524}, no.4, 6167-6180 (2023)
doi:10.1093/mnras/stad2262
[arXiv:2212.04521 [astro-ph.CO]].

\bibitem{vonWietersheim-Kramsta:2024cks}
M.~von Wietersheim-Kramsta, K.~Lin, N.~Tessore, B.~Joachimi, A.~Loureiro, R.~Reischke and A.~H.~Wright,
Astron. Astrophys. \textbf{694}, A223 (2025)
doi:10.1051/0004-6361/202450487
[arXiv:2404.15402 [astro-ph.CO]].

\bibitem{DES:2024jgw}
M.~Gatti \textit{et al.} [DES],
Phys. Rev. D \textbf{111}, no.6, 063504 (2025)
doi:10.1103/PhysRevD.111.063504
[arXiv:2405.10881 [astro-ph.CO]].

\bibitem{Novaes:2024dyh}
C.~P.~Novaes, L.~Thiele, J.~Armijo, S.~Cheng, J.~A.~Cowell, G.~A.~Marques, E.~G.~M.~Ferreira, M.~Shirasaki, K.~Osato and J.~Liu,
Phys. Rev. D \textbf{111}, no.8, 083510 (2025)
doi:10.1103/PhysRevD.111.083510
[arXiv:2409.01301 [astro-ph.CO]].

\bibitem{Homer:2024cwg}
J.~Homer, O.~Friedrich and D.~Gruen,
Astron. Astrophys. \textbf{699}, A213 (2025)
doi:10.1051/0004-6361/202453339
[arXiv:2412.02311 [astro-ph.CO]].

\bibitem{Saoulis:2025bug}
A.~A.~Saoulis, D.~Piras, N.~Jeffrey, A.~Spurio Mancini, A.~M.~G.~Ferreira and B.~Joachimi,
Mon. Not. Roy. Astron. Soc. \textbf{542}, no.4, 3231-3245 (2025)
doi:10.1093/mnras/staf1436
[arXiv:2505.21215 [astro-ph.CO]].

\bibitem{Diao:2025szg}
K.~Diao, B.~Dai and U.~Seljak,
JCAP \textbf{08}, 004 (2025)
doi:10.1088/1475-7516/2025/08/004
[arXiv:2505.00632 [astro-ph.CO]].

\bibitem{Shirasaki:2018thk}
M.~Shirasaki, N.~Yoshida and S.~Ikeda,
Phys. Rev. D \textbf{100}, no.4, 043527 (2019)
doi:10.1103/PhysRevD.100.043527
[arXiv:1812.05781 [astro-ph.CO]].

\bibitem{Jeffrey:2019fag}
N.~Jeffrey, F.~Lanusse, O.~Lahav and J.~L.~Starck,
Mon. Not. Roy. Astron. Soc. \textbf{492}, no.4, 5023-5029 (2020)
doi:10.1093/mnras/staa127
[arXiv:1908.00543 [astro-ph.CO]].

\bibitem{Remy:2022ixn}
B.~Remy, F.~Lanusse, N.~Jeffrey, J.~Liu, J.~L.~Starck, K.~Osato and T.~Schrabback,
Astron. Astrophys. \textbf{672}, A51 (2023)
doi:10.1051/0004-6361/202243054
[arXiv:2201.05561 [astro-ph.CO]].

\bibitem{Whitney:2024pmf}
J.~J.~Whitney, T.~I.~Liaudat, M.~A.~Price, M.~Mars and J.~D.~McEwen,
Mon. Not. Roy. Astron. Soc. \textbf{542}, no.3, 2464-2479 (2025)
doi:10.1093/mnras/staf1356
[arXiv:2410.24197 [astro-ph.CO]].

\bibitem{Aoyama:2025aut}
S.~D.~Aoyama, K.~Osato and M.~Shirasaki,
[arXiv:2505.00345 [astro-ph.CO]].

\bibitem{Wang:2026ygy}
Y.~Wang and Y.~Yu,
Phys. Rev. D \textbf{113}, 043553 (2026)
doi:10.1103/kc9z-jllp
[arXiv:2603.25471 [astro-ph.IM]].

\bibitem{Mustafa:2017bqp}
M.~Mustafa, D.~Bard, W.~Bhimji, Z.~Luki{\'c}, R.~Al-Rfou and J.~M.~Kratochvil,
Comput. Astrophys. Cosmol. \textbf{6}, no.1, 1 (2019)
doi:10.1186/s40668-019-0029-9
[arXiv:1706.02390 [astro-ph.IM]].

\bibitem{Perraudin:2020gig}
N.~Perraudin, S.~Marcon, A.~Lucchi and T.~Kacprzak,
Front. Artif. Intell. \textbf{4}, 673062 (2021)
doi:10.3389/frai.2021.673062
[arXiv:2004.08139 [astro-ph.CO]].

\bibitem{Tamosiunas:2020rvw}
A.~Tamosiunas, H.~A.~Winther, K.~Koyama, D.~J.~Bacon, R.~C.~Nichol and B.~Mawdsley,
Mon. Not. Roy. Astron. Soc. \textbf{506}, no.2, 3049-3067 (2021)
doi:10.1093/mnras/stab1879
[arXiv:2004.10223 [astro-ph.CO]].

\bibitem{Shirasaki:2023nnk}
M.~Shirasaki and S.~Ikeda,
doi:10.33232/001c.118104
[arXiv:2310.17141 [astro-ph.CO]].

\bibitem{Boruah:2024rgr}
S.~S.~Boruah, P.~Fiedorowicz, R.~Garcia, W.~R.~Coulton, E.~Rozo and G.~Fabbian,
[arXiv:2406.05867 [astro-ph.CO]].

\bibitem{Boruah:2025zkn}
S.~S.~Boruah, M.~Jacob and B.~Jain,
Phys. Rev. D \textbf{111}, no.8, 083542 (2025)
doi:10.1103/PhysRevD.111.083542
[arXiv:2502.04158 [astro-ph.CO]].

\bibitem{Boruah:2025iwf}
S.~S.~Boruah, M.~Jacob, B.~Jain, R.~Maiya and R.~Venkataramanan,
[arXiv:2511.14667 [astro-ph.CO]].

\bibitem{Dai:2022dso}
B.~Dai and U.~Seljak,
Mon. Not. Roy. Astron. Soc. \textbf{516}, no.2, 2363-2373 (2022)
doi:10.1093/mnras/stac2010
[arXiv:2202.05282 [astro-ph.CO]].

\bibitem{Armijo:2026ucb}
J.~Armijo, L.~Thiele and J.~Liu,
[arXiv:2601.20669 [astro-ph.CO]].

\bibitem{Fiedorowicz:2021clg}
P.~Fiedorowicz, E.~Rozo, S.~S.~Boruah, C.~Chang and M.~Gatti,
Mon. Not. Roy. Astron. Soc. \textbf{512}, no.1, 73-85 (2022)
doi:10.1093/mnras/stac468
[arXiv:2105.14699 [astro-ph.CO]].

\bibitem{Boruah:2022lsu}
S.~S.~Boruah, E.~Rozo and P.~Fiedorowicz,
Mon. Not. Roy. Astron. Soc. \textbf{516}, no.3, 4111-4122 (2022)
doi:10.1093/mnras/stac2508
[arXiv:2204.13216 [astro-ph.CO]].

\bibitem{Zhou:2023ezg}
A.~J.~Zhou, X.~Li, S.~Dodelson and R.~Mandelbaum,
Phys. Rev. D \textbf{110}, no.2, 023539 (2024)
doi:10.1103/PhysRevD.110.023539
[arXiv:2312.08934 [astro-ph.CO]].

\bibitem{Boruah:2024tqp}
S.~S.~Boruah, P.~Fiedorowicz and E.~Rozo,
Phys. Rev. D \textbf{110}, no.2, 023524 (2024)
doi:10.1103/PhysRevD.110.023524
[arXiv:2403.05484 [astro-ph.CO]].

\bibitem{Porqueres:2023drp}
N.~Porqueres, A.~Heavens, D.~Mortlock, G.~Lavaux and T.~L.~Makinen,
[arXiv:2304.04785 [astro-ph.CO]].

\bibitem{Massey:2007wb}
R.~Massey, J.~Rhodes, R.~Ellis, N.~Scoville, A.~Leauthaud, A.~Finoguenov, P.~Capak, D.~Bacon, H.~Aussel and J.~P.~Kneib, \textit{et al.}
Nature \textbf{445}, 286 (2007)
doi:10.1038/nature05497
[arXiv:astro-ph/0701594 [astro-ph]].

\bibitem{Oguri:2017vrv}
M.~Oguri, S.~Miyazaki, C.~Hikage, R.~Mandelbaum, Y.~Utsumi, H.~Miyatake, M.~Takada, R.~Armstrong, J.~Bosch and Y.~Komiyama, \textit{et al.}
Publ. Astron. Soc. Jap. \textbf{70}, S26 (2018)
doi:10.1093/pasj/psx070
[arXiv:1705.06792 [astro-ph.CO]].

\bibitem{Leonard:2013hia}
A.~Leonard, F.~Lanusse and J.~L.~Starck,
Mon. Not. Roy. Astron. Soc. \textbf{440}, no.2, 1281-1294 (2014)
doi:10.1093/mnras/stu273
[arXiv:1308.1353 [astro-ph.CO]].

\bibitem{Leonard:2015jha}
A.~Leonard, F.~Lanusse and J.~L.~Starck,
Mon. Not. Roy. Astron. Soc. \textbf{449}, no.1, 1146-1157 (2015)
doi:10.1093/mnras/stv386
[arXiv:1502.05872 [astro-ph.CO]].

\bibitem{Li:2021vsa}
X.~Li, N.~Yoshida, M.~Oguri, S.~Ikeda and W.~Luo,
Astrophys. J. \textbf{916}, no.2, 67 (2021)
doi:10.3847/1538-4357/ac0625
[arXiv:2102.09707 [astro-ph.CO]].

\end{thebibliography}
\end{document}